\label{key}
\documentclass[a4paper,fleqn,usenatbib]{mnras}

\usepackage{newtxtext,newtxmath}

\usepackage[T1]{fontenc}
\usepackage{ae,aecompl}

\usepackage{graphicx}	
\graphicspath{{fig/}}
\usepackage{siunitx}
\DeclareSIUnit\gauss{G}
\DeclareSIUnit\erg{erg}
\DeclareSIUnit\parsec{pc}
\usepackage{amsmath}	
\usepackage{amssymb}	
\usepackage{physics}
\usepackage{todonotes}
\presetkeys{todonotes}{inline, size=\tiny}{}
\usepackage[normalem]{ulem}
\usepackage{xcolor}
\usepackage{xparse}

\newcommand{\mat}{\ensuremath{\mathbfss}}
\newcommand{\bs}[1]{\boldsymbol{#1}}
\newcommand{\e}{\mathrm{e}}
\renewcommand{\vec}{\ensuremath{\mathbfit}}
\newcommand{\mo}{\ensuremath{\mathopen{}}}
\newcommand{\ratio}{\ensuremath{x}}

\title[Evolution of cosmic ray electron spectra]{Evolution of cosmic ray electron spectra in magnetohydrodynamical simulations}

\author[G. Winner et al.]{
Georg Winner,$^{1, 2}$\thanks{E-mail: gwinner@aip.de (GW)}
Christoph Pfrommer,$^{1}$
Philipp Girichidis,$^{1}$
R{\"u}diger Pakmor$^{3}$
\\
$^{1}$Leibniz-Institut f{\"u}r Astrophysik Potsdam (AIP), An der Sternwarte 16, 14482 Potsdam, Germany\\
$^{2}$Fakult{\"a}t f{\"u}r Physik und Astronomie, Universit{\"a}t Heidelberg, Im Neuenheimer Feld 226, 69120  Heidelberg, Germany\\
$^{3}$Max-Planck-Institut f{\"u}r Astrophysik, Karl-Schwarzschild-Str. 1, 85741 Garching, Germany\\
}

\date{Accepted 2019 June 27. Received 2019 June 19; in original form 2019 March 4}

\pubyear{2019}

\begin{document}
\label{firstpage}
\pagerange{\pageref{firstpage}--\pageref{lastpage}}
\maketitle

\begin{abstract}
Cosmic ray (CR) electrons reveal key insights into the non-thermal physics of
the interstellar medium (ISM), galaxies, galaxy clusters, and active galactic nuclei
by means of their inverse Compton $\gamma$-ray emission and synchrotron emission
in magnetic fields. While magnetohydrodynamical (MHD) simulations with CR
protons capture their dynamical impact on these systems, only few computational
studies include CR electron physics because of the short cooling time-scales and
complex hysteresis effects, which require a numerically expensive,
high-resolution spectral treatment. Since CR electrons produce important
non-thermal observational signatures, such a spectral CR electron treatment is
important to link MHD simulations to observations. We present an efficient
post-processing code for Cosmic Ray Electron Spectra that are evolved in Time
(\textsc{crest}) on Lagrangian tracer particles. The CR electron spectra are
very accurately evolved on comparably large MHD time steps owing to an
innovative hybrid numerical-analytical scheme. \textsc{crest} is coupled to the
cosmological MHD code \textsc{arepo} and treats all important aspects of
spectral CR electron evolution such as adiabatic expansion and compression,
Coulomb losses, radiative losses in form of inverse Compton, bremsstrahlung and
synchrotron processes, diffusive shock acceleration and reacceleration,
Fermi-II reacceleration, and secondary electron injection. After showing various code validations of idealized
one-zone simulations, we study the coupling of \textsc{crest} to MHD
simulations. We demonstrate that the CR electron spectra are efficiently and
accurately evolved in shock-tube and Sedov--Taylor blast wave simulations. This
opens up the possibility to produce self-consistent synthetic observables of
non-thermal emission processes in various astrophysical environments.
\end{abstract}




\section{Introduction}
CRs are ubiquitous in many astrophysical environments, such as the ISM, galaxies, galaxy clusters and active galactic nuclei (AGN). CRs are
non-thermal, charged particles consisting of a hadronic component (mainly
protons and alpha particles) as well as leptons (mainly electrons and
positrons). The leptonic component (henceforth referred to as CR electrons)
suffers fast radiative losses via synchrotron interactions with magnetic fields
and inverse Compton (IC) interactions with ambient photon fields. Hence they are
directly linked to observations of the non-thermal emission from radio to
gamma-ray wavelengths. Hadronic CRs (henceforth referred to as CR protons) are
interesting since they play an important dynamical role in the ISM due to their energy equipartition with turbulent and magnetic energy in
the midplane of the Milky Way \citep{Boulares1990}. As CR protons stream and
diffuse vertically from their sources in the galactic midplane, their emerging
CR proton pressure gradient can dominate the force balance and accelerate the
gas, thus driving a galactic outflow as shown in one-dimensional (1D) magnetic
flux-tube models \citep{Breitschwerdt1991, Zirakashvili1996, Ptuskin1997,
Everett2008, Samui2018} and three-dimensional (3D) simulations \citep{Uhlig2012,
Booth2013, Salem2014, Pakmor2016a,Simpson2016, Girichidis2016, Pfrommer2017b,
Ruszkowski2017, Jacob2018}.

Fast-streaming CR protons resonantly excite Alfv\'en waves through the ``streaming
instability'' \citep{Kulsrud1969}. Damping of these waves effectively
transfers CR to thermal energy. This process is thought to provide the physical
heating mechanism underlying the ``cooling flow problem'' in galaxy clusters
where the cooling gas and nuclear activity appear to be tightly coupled to a
self-regulated feedback loop \citep{McNamara2007}. As CR protons stream out of AGN
lobes they can stably heat the surrounding cooling intracluster medium
\citep{Loewenstein1991, Guo2008, Ensslin2011, Fujita2012, Pfrommer2013,
Jacob2017, Jacob2017a, Ruszkowski2017a, Ehlert2018}.

Early studies of CR protons in computational cosmology were performed by the Eulerian
mesh \textsc{cosmocr} code \citep{Miniati2001} and at cosmological shocks by
N-body/hydrodynamical simulations \citep{Ryu2003}. The first MHD simulations
with active CR proton transport were performed with the \textsc{zeus-3d}
code \citep{Hanasz2003}.  Modeling CR proton physics in the smoothed particle
hydrodynamics code \textsc{gadget-2} enabled adaptive spatial resolution in
high-density environments and to explore the impact of CR protons on the formation of
galaxies and galaxy clusters
\citep{Pfrommer2006,Ensslin2007,Jubelgas2008}. Further numerical CR proton studies
were performed with the Eulerian mesh code \textsc{piernik} \citep{Hanasz2010},
the adaptive mesh refinement codes \textsc{ramses} \citep{Booth2013,
Dubois2016}, \textsc{enzo} \citep{Salem2014}, \textsc{flash} \citep{Girichidis2016,
Girichidis2018}, and \textsc{pluto} \citep{Mignone2018}, the moving-mesh
code \textsc{arepo} \citep{Pakmor2016b,Pfrommer2017a}.

In comparison to CR protons, the energy of CR electrons falls short by a factor
of about 100 at the solar radius in the Milky Way \citep{Zweibel2013}; hence CR
electrons are not dynamically important. The cooling time-scale of
relativistic CR electrons with Lorentz factors $\gamma\gtrsim10^3$ is much
shorter than that of relativistic CR protons at the same energy per
particle. While CR protons can only effectively cool via rare hadronic
interactions (thereby lowering the resulting luminosity), CR electrons cool
efficiently via synchrotron and IC interactions. This means that much of the
non-thermal physics is only observationally accessible through the leptonic
emission channel. Thus, it is very important to model the momentum spectrum of
CR electrons alongside \mbox{(magneto)}hydrodynamical simulations in order to
produce realistic synthetic non-thermal observables. Comparing those to
observational data enables scrutinising our simulated physics and our
understanding of galaxy formation, evolution of galaxy clusters or AGN jet
physics. 

Supernova remnants (SNRs) provide us with important insights into the physics of
particle acceleration and have been observed from radio to $\gamma$-ray
energies \citep{Helder2012, Blasi2013, Bykov2018}. This radiation is produced by
hadronic and leptonic processes, and the ambient density and the magnetic field
strength of the SNR determine which of these processes dominates. In low-density
environments of SNR, such as RX~J1713.7-3946, IC emission by CR electrons likely
dominates the $\gamma$-ray emission (\citealt{Ellison2012}, but
see \citealt{Celli2019a}, for an interpretation in terms of hadronic
emission). Stellar bow shocks of massive runaway stars are also a site of
particle acceleration, e.g. the radio emission observed in the bow shock of the
runaway star BD~+$43^\circ3654$ might be produced by synchrotron radiation of
CR electrons \citep{Benaglia2010}.

Many galaxies exhibit galactic outflows that shine in radio, X-rays, and
$\gamma$-rays.  Understanding the physics of galactic outflows is the holy grail
of galaxy formation. The most prominent example of these outflows are the Fermi
bubbles, which extend to about \SI{8}{kpc} north and south from the central
region of our Milky Way. They are observed as hard-spectrum gamma-ray
structures \citep{Su2010, Dobler2010} which coincide with radio
lobes \citep{Caretti2013}. The origin of the Fermi bubbles remains elusive and
it is not clear whether hadronic CR proton interactions or leptonic IC emission
scenarios are dominant for the observed $\gamma$-ray emission.  Models generally
rely on AGN or starburst events. There are several attempts to simulate the
evolution of the Fermi bubbles \citep{Yang2017, Mertsch2019} or more generally,
to understand radio signatures of outflows in external
galaxies \citep{Heesen2016}. However,
self-consistent \mbox{(magneto)}hydrodynamical simulations of the Milky Way
with CR proton and electron physics are still missing.

Galaxy clusters shine in radio due to synchrotron emission of CR electrons in turbulent
cluster magnetic fields. There are three important classes of radio sources in
galaxy clusters: radio relics, giant haloes and radio mini
haloes \citep{Bykov2019}.  Giant radio haloes are characterised by spatially
extended regions of diffuse, unpolarised radio emission with an irregular
morphology that is centred on the cluster. In contrast, radio relics are often
located at the periphery of clusters and show a high degree of polarization with an
irregular, elongated morphology. There are several simulation studies of CR electron
acceleration and diffuse radio synchrotron emission in the context of galaxy
clusters \citep[e.g.][]{Miniati2001a, Miniati2003, Pfrommer2008, Battaglia2009,
Pinzke2013, Pinzke2017, Vazza2012, Donnert2013, Donnert2014, Guo2014a, Guo2014b,
Kang2019}.

The plethora of astrophysical systems that shine through leptonic non-thermal
radiation makes it inevitable to evolve the CR proton and electron physics on
top of MHD simulations in order to distinguish hadronic and leptonic scenarios.
Despite the importance of CR electrons, there are only few numerical codes that
can evolve the spectra of CR electrons in MHD simulations, e.g.
the \textsc{pluto} code with CR electrons on Lagrangian
particles \citep{Vaidya2018}. We aim at further closing this gap by presenting
a numerical post-processing code for Cosmic Ray Electron Spectra that are
evolved in Time (\textsc{crest})\footnote{The name \textsc{crest} also refers to
the physical phenomenon of CR electrons being accelerated and swept up by a
shock wave while shining on its crest via synchrotron and IC radiation.}, which
works together with \mbox{(magneto)}hydrodynamical codes that have
Lagrangian tracer particles. In this work, we present the algorithm and test its
implementation in one-zone problems. To evolve the CR electron spectrum
spatially and temporally resolved alongside MHD simulations, we
couple \textsc{crest} to the massively-parallel hydrodynamical
code \textsc{arepo} \citep{Springel2010}, that can also follow CR proton
physics \citep{Pfrommer2017a}.  In evolving the CR electron
spectrum, \textsc{crest} includes adiabatic effects, all important energy loss
processes of CR electron as well as energy gain processes such as diffusive
shock acceleration (via the Fermi-I process) and reacceleration, Fermi-II
reacceleration via particle interactions with compressible turbulence, and
secondary electron injection.

We present the physical and numerical foundations of our algorithm in
section~\ref{sec:Method}. We proceed with numerical tests of our code, including
idealized one-zone tests in section~\ref{sec:Idealized_One_Zone_Tests} and
simulations with \textsc{arepo} in section~\ref{sec:Simulations_with_arepo}.  We
conclude in section~\ref{sec:Conclusion} and provide an outlook of future
astrophysical applications of our work. In appendix~\ref{sec:AppA}, we detail
the discretisation scheme and numerical algorithms adopted for solving the
Fokker--Planck equation of CR electrons. We use the cgs system of units
throughout this work.

\section{Methodology}
\label{sec:Method}

Here, we introduce the theoretical background before we explain our
discretisation scheme and numerical algorithms to describe our subgrid scale
model for Fermi-I acceleration. We then present analytical solutions of limiting
cases and our hybrid algorithm that combines analytical and numerical solutions
to the transport equation of CR electrons.

\subsection{Theoretical background}
\subsubsection{Transport equation}
The CR electron distribution is completely described by the phase space density
$f(\vb*{x},\vb*{p},t)$ whose evolution is given by the relativistic Vlasov
equation. Throughout this paper, we use the dimensionless electron momentum,
$\vb*{p}=\vb*{P}/(m_\e c)$.  CR electrons gyrate around magnetic field lines
which are subject to random fluctuations. The application of quasi-linear theory
by ensemble averaging over fluctuations, and the use of the diffusion
approximation, i.e. the assumption of near-isotropic equilibrium as a
consequence of frequent pitch-angle scattering on MHD turbulence leads to the
Fokker--Planck equation \citep{Schlickeiser1989a, Zank2014eltit}.

We follow the transport of CR electrons on Lagrangian tracer particles and
include continuous losses plus a source term \citep{Schlickeiser1989b}.  Here,
we assume that CR electrons are transported with the gas as they are confined to
their gyration orbits around turbulent magnetic fields, which are frozen into
the moving plasma. The Fokker--Planck equation for the 1D distribution in
momentum space is related to the 3D distribution via $f(p) = 4\pi p^2
f^{\rmn{3D}}(p)$ and obeys the Fokker--Planck equation without CR
streaming \citep[e.g.][]{Pinzke2017}
\begin{equation}
\dv{f(p,t)}{t} = 
	\begin{aligned}[t]
		& \phantom{{}+{}} \pdv{p} \left[ f(p,t) \frac{p}{3} \left(\div{\vb*{\varv}}\right)\right]  - \left(\div{\vb*{\varv}}\right) f(p,t) \\
		& - \pdv{p} \left[ f(p,t) \dot{p}(p,t) \right] + Q(p,t) \\
		& - \pdv{p} \left[ \frac{f(p,t)}{p^2} \pdv{p}(p^2 D_{\!p\!p}) \right] + \pdv[2]{p}\left[D_{\!p\!p} f(p,t)\right]\\
		& + \bs{\nabla}  \bs{\cdot} \left[\mat{K} \bs{\cdot} \bs{\nabla} f(p, t) \right],
	\end{aligned} \label{eq:CRe_1D_FP_full}
\end{equation}
where $\dv*{}{t} = \upartial/\upartial t + \bs{\varv\cdot\nabla}$ is the
Lagrangian time derivative and $p = \abs{\vb*{p}}$ is the absolute value of the
momentum.  The first line on the right-hand side describes adiabatic changes
resulting from changes in the gas velocity~$\vb*{\varv}$ and Fermi-I
acceleration and reacceleration \citep[in combination with spatial diffusion,
see][]{Blandford1987}.

The second line describes energy losses (i.e. Coulomb and radiative losses)
$\dot{p}(p, t)$ and injection with source function $Q(p,t)$ for unresolved
subgrid acceleration processes and secondary electron injection that are
produced in hadronic interactions of CR protons with the ambient gas. The latter
process is described by $Q_\mathrm{inj} = \dot{C}_\mathrm{inj}
p^{-\alpha_\mathrm{inj}}$ with injection slope $\alpha_\mathrm{inj}$ that is
identical to that of the CR proton distribution, an injection rate
$\dot{C}_\mathrm{inj}=C_\mathrm{inj}/\tau_\rmn{pp}$, where
$C_\mathrm{inj}\propto n_\mathrm{crp}$ and $\tau_\rmn{pp}=
1/(c \sigma_\mathrm{pp} n_\mathrm{tar})$ \citep{Mannheim1994}. Here, $c$ is the
speed of light, $\sigma_\mathrm{pp}$ is the proton-proton cross-section,
$n_\mathrm{tar}$ is the target proton density, and $n_\mathrm{crp}$ is the
number density of CR protons, which we dynamically evolve with the CR proton
module of \textsc{arepo} \citep{Pfrommer2017a}.

The third line represents the momentum diffusion (Fermi-II reacceleration) with
a momentum-dependent diffusion $D_{\!p\!p}(p)$ while the last line describes
spatial CR diffusion with the diffusion tensor $\mat{K}$. Because we do not
resolve the necessary scales and plasma processes to directly follow diffusive
shock acceleration via the adiabatic and diffusive terms, we have to treat
Fermi-I acceleration and reacceleration in form of an analytic subgrid model via
the source term $Q(p, t)$ in our code. We defer the explicit treatment of
spatial CR diffusion, as well as CR streaming, to future studies.

\subsubsection{Loss processes}
We note that energy losses are in general time dependent as photon fields,
magnetic fields and electron number densities change in time. We will suppress
the explicit time dependence in the following formulae for simplicity. Coulomb
losses \citep{Gould1972} are described by
\begin{align}
\dot{p}_\mathrm{c}(p) = - \frac{3 \sigma_\mathrm{T} n_\e c}{2\beta^2} 
\begin{aligned}[t] 
& \left[\ln( \frac{m_\e c^2 \beta \sqrt{\gamma - 1}}{\hbar \omega_\mathrm{pl}} ) \vphantom{\left( \frac{\gamma - 1}{4 \gamma}\right)^2} \right.\\
& \left.\ + \ln(2) \left(\frac{\beta^2}{2} + \frac{1}{\gamma} \right) +\frac{1}{2} + \left( \frac{\gamma - 1}{4 \gamma}\right)^2  \vphantom{\ln( \frac{m_\e c^2 \beta \sqrt{\gamma - 1}}{\hbar \omega_\mathrm{pl}} )} \right],
\end{aligned} \label{eq:Coulomb_loss_rate_exact}
\end{align}
where $\sigma_\mathrm{T}=8 \pi e^4 ( m_\mathrm{e} c^2)^{-2} / 3 $ is the Thomson
cross-section, $\hbar$ is the reduced Planck constant, $m_\mathrm{e}$ the electron mass, $\beta = p (1+p^2)^{-1/2}$ is the
dimensionless CR electron velocity, and $\gamma = (1+p^2)^{1/2}$ is the Lorentz
factor of CR electrons.  The electron density is $n_\e = n_\mathrm{gas}
X_\mathrm{H} x_\e$ where $X_\mathrm{H}$ is the hydrogen mass fraction and
$x_\e=n_\e / n_\mathrm{H}$ is the ionization fraction, the ratio of electron
density-to-hydrogen density, which is denoted by $n_\mathrm{H}$. The plasma
frequency is $\omega_\mathrm{pl} = \sqrt{4\pi e^2 n_\e / m_\e}$ and $e$ denotes
the elementary charge.

Charged particles experience synchrotron losses in magnetic fields and
experience inverse Compton scattering off of photon
fields \citep{Rybicki1986}. Synchrotron losses are given by
\begin{align}
\dot{p}_\mathrm{s}(p) =  - \frac{4 \sigma_\mathrm{T} p^2}{3 m_\e c \beta} \frac{B^2}{8\pi}, \label{eq:Synchrotron_loss_rate_exact}
\end{align}
and inverse Compton processes by
\begin{align}
\dot{p}_\mathrm{ic}(p) = - \frac{4 \sigma_\mathrm{T} p^2}{3 m_\e c \beta} \varepsilon_\rmn{ph},
\label{eq:Inverse_Compton_loss_rate_exact}
\end{align}
where the total radiation field is a sum over the cosmic microwave background
(CMB) radiation and star light, $\varepsilon_\rmn{ph} = \varepsilon_\mathrm{star}
+ \varepsilon_\mathrm{cmb}$.  The momentum loss rate of the bremsstrahlung loss
process is given by
\begin{align}
\dot{p}_\mathrm{b}(p) = - \frac{16}{3} \alpha \left(\frac{e^2}{m_\e^2 c^3}\right)^2 \gamma \chi\left[E(p)\right],
\end{align}
where $\alpha$ is the fine-structure constant and the function $\chi\left[E(p)\right]$ is
provided by \cite{Koch1959}. The total energy loss rate is given by the sum of
all losses:
\begin{align}
\dot{p} (p,t) = \dot{p}_\rmn{c} (p) + \dot{p}_\rmn{s} (p) + \dot{p}_\rmn{ic} (p) + \dot{p}_\rmn{b} (p).
\label{eq:pdot}
\end{align}

\subsubsection{Fermi-I acceleration and reacceleration}
Diffusive shock acceleration also known as Fermi-I acceleration is an important
energy gain process for CR electrons. It is a combination of direct acceleration
of electrons from the thermal pool and of reacceleration of a fossil electron
distribution $f_\mathrm{pre}$ in the pre-shock region, if present.

The total spectrum in the post-shock region is obtained by evaluating adiabatic
changes and spatial diffusion of equation~\eqref{eq:CRe_1D_FP_full} at the shock. The
analytic solution of the total post-shock spectrum is \citep{Bell1978b,
Drury1983, Blandford1987}
\begin{align}
f_\mathrm{post} (p) = f_\mathrm{reac}(p) + f_\mathrm{acc}(p),
\label{eq:Fermi_I_PostShockSpectrum}
\end{align}
where the reaccelerated and accelerated spectrum are
\begin{align}
f_\mathrm{reac}(p) &= (\alpha_\mathrm{acc} + 2) p^{-\alpha_\mathrm{acc}} \int_{p_\mathrm{inj}}^{p} p'^{\alpha_\mathrm{acc} - 1} f_\mathrm{pre}(p') \dd{p'} \mbox{ and} \label{eq:Fermi_I_reacceleration_spectrum}\\
f_\mathrm{acc}(p) &= C_\mathrm{acc} p^{-\alpha_\mathrm{acc}} \Theta(p - p_\mathrm{inj}), \label{eq:Fermi_I_acceleration_spectrum}
\end{align}
respectively, where $C_\mathrm{acc}$ is the normalization and $p_\mathrm{inj}$
is the injection momentum of the accelerated spectrum.  The spectral index
$\alpha_\mathrm{acc}$ is calculated by
\begin{align}
\alpha_\mathrm{acc} = \frac{r + 2}{r - 1},
\end{align}
where $r=\rho_\mathrm{post}/\rho_\mathrm{pre}$ denotes the shock compression
ratio, i.e. the ratio of post-shock to pre-shock gas density. We also take 
cooling processes into account, which lead to a modified spectrum with a
momentum cutoff \citep{Ensslin1998, Zirakashvili2007, Pinzke2010} of the form
\begin{align}
\tilde{f}_\mathrm{post}(p) = f_\mathrm{post}(p) \left[1 + a \left(\frac{p}{p_\mathrm{acc}}\right)^b \right]^c \exp[ - \left(\frac{p}{p_\mathrm{acc}}\right)^2],
\end{align}
where we adopt the parameters $a=0.66$, $b=2.5$, and $c=1.8$ and
$p_\mathrm{acc}$ is the cutoff momentum of the (re)accelerated spectrum
\begin{align}
p_\mathrm{acc} = \begin{aligned}[t] 
& \frac{\varv_\rmn{post}}{c} \sqrt{\frac{3 e (r - 1)}{4 \sigma_\mathrm{T}}}\\
& \times \left[
  \frac{B_\mathrm{pre}^2/(8 \pi) + \varepsilon_\rmn{ph}}{r B_\mathrm{pre}} + \frac{B_\mathrm{post}^2/(8 \pi) + \varepsilon_\rmn{ph}}{B_\mathrm{post}}    \right]^{-1/2},
  \end{aligned}
\end{align}
where $\varv_\rmn{post}$ is the post-shock velocity in the shock rest
frame and $B_\mathrm{pre}$ and $B_\mathrm{post}$ are the pre- and post-shock magnetic fields. Here, we assume a parallel shock geometry so that the magnetic field
strength is constant across the shock. We postpone a modelling of the
dependencies of the maximum electron energy on magnetic obliquity and amplified
magnetic fields via plasma effects such as the non-resonant hybrid instability
driven by the CR proton current propagating upstream of the
shock \citep{Bell2004}.

\subsubsection{Fermi-II reacceleration}
Stochastic acceleration, originally proposed by \cite{Fermi1949}, describes the
energy gains of CRs through random collisions with plasma waves and
turbulence. As the gain per collision process is of second order in the velocity
ratio of collision counterpart to particle, it is also referred to as Fermi-II
reacceleration \citep{Petrosian2012}.  However, Coulomb cooling is too fast for
stochastic acceleration from the thermal pool to be efficient in cluster and
galactic environments \citep{Petrosian2001}. Therefore, the Fermi-II process is
only efficient in reaccelerating a fossil non-thermal electron distribution.

Fermi-II reacceleration by turbulent magnetic fields was investigated in galaxy
clusters as primary energy source for diffusive radio emission from CR electrons
in the Coma cluster \citep{Jaffe1977, Schlickeiser1987}. There are different
energy transfer channels of turbulent energy injection into CR,
e.g. via magnetosonic waves \citep{Ptuskin1988} or
via transit time damping (TTD) of compressible fast magnetosonic
modes \citep{Brunetti2007, Brunetti2011}.

CRs gain energy in turbulent reacceleration through transit time damping. The
momentum diffusion in equation~\eqref{eq:CRe_1D_FP_full} is given by
\begin{align}
D_{\!p\!p} = D_0 p^2 \label{eq:MomentumDiffusion_Function}
\end{align}
where the physics of turbulent reacceleration is encapsulated in the constant
$D_0$ \citep{Pinzke2017}. The momentum diffusion time is $\tau_{\!p\!p} = p^2 /
(4 D_{\!p\!p})$ which is $\tau_{\!p\!p} = 1/(4 D_0)$ according to equation~\eqref{eq:MomentumDiffusion_Function}.

\subsection{Numerical discretisation}
\label{sec:Numerical_discretisation}

\subsubsection{General setup}

In order to solve equation~\eqref{eq:CRe_1D_FP_full} numerically, we apply three
discretisations to the CR electron phase space density $f=f(\vb*{x},p,t)$.  (i)
We discretise $f$ in configuration space with Lagrangian tracer particles, (ii)
we discretise the momentum spectrum of every tracer particle with piecewise
constant values per momentum bin, and (iii) $f$ is discretised in time.  The
momentum grid is equally spaced in logarithmic space and we use $N$ bins between
the lowest momentum $p_\mathrm{min}$ and highest momentum $p_\mathrm{max}$. The
bin centres are located at
\begin{align}
&p_i = p_\mathrm{min} \exp[ \left(i+\frac{1}{2}\right) \Delta \ln p] &&\mbox{for } i=0, 1, \ldots, N - 1, \label{eq:BinCenters}
\end{align}
and the bin edges are given by
\begin{align}
&p_{i - \frac{1}{2}} = p_\mathrm{min} \exp( i\, \Delta \ln p) &&\mbox{for } i=0, 1, \ldots, N, \label{eq:BinEdges}
\end{align}
where $\Delta \ln p = \ln(p_\mathrm{max} / p_\mathrm{min})/N$ is the grid
spacing.  The spectrum is defined on all bin centers and is evolved in time from
$t$ by a time step $\Delta t$ with an operator split approach,
\begin{align}
f(\vb*{x},p, t + \Delta t) = \mathcal{A}_\mathrm{diff}\left(\frac{\Delta t}{2}\right) \mathcal{A}_\mathrm{adv}\left(\Delta t\right) \mathcal{A}_\mathrm{diff}\left(\frac{\Delta t}{2}\right) f(\vb*{x},p, t). \label{eq:EvolutionByOperatorSplit}
\end{align}
Adiabatic changes, Fermi-I (re)acceleration, cooling, and injection are
calculated with an advection operator $\mathcal{A}_\mathrm{adv}$ and diffusion
in momentum space is calculated with a diffusion operator
$\mathcal{A}_\mathrm{diff}$ that both advance the solution for the time step of
their arguments.

The advection operator is based on a flux-conserving finite volume scheme with a
second-order piecewise linear reconstruction of the spectrum.  The terms
$\pdv*{p}\left\{ f(p) [p ( \div{\vb*{\varv}} ) / 3 - \dot{p}]\right\}$, which
include cooling and partially adiabatic changes, are interpreted as advection in
momentum space in order to calculate fluxes across the bin edges given in
equation~\eqref{eq:BinEdges}.  In addition, we use the non-linear van Leer flux
limiter \citep{Leer1977}.  The remaining terms for injection,
Fermi-I \mbox{(re)acceleration} and adiabatic changes, are treated as an
inhomogeneity of the partial differential equation (for details, see
appendix~\ref{sec:advection_op}). Our implementation is second-order accurate in
time and momentum space.

The diffusion operator is based on a finite difference scheme with a
semi-implicit Crank--Nicolson algorithm, which is accurate to second order in
time and to first order in momentum space (for details, see
appendix \ref{sec:diffusion_op}).

\subsubsection{Time steps and characteristic momenta}

The overall time step $\Delta t$ in equation~\eqref{eq:EvolutionByOperatorSplit} is determined by
\begin{align}
\Delta t = \min\mo\left(\Delta t_\mathrm{adv}, \Delta t_\mathrm{diff} \right),
\end{align}
the minimum of the time step for advection and diffusion,
\begin{align}
&\Delta t_\mathrm{adv} = C_{\rmn{CFL}} \left[ \max\mo\left(\frac{\abs{\dot{p}(p)}}{\Delta p}\right) + \frac{\abs{\Delta n}}{n} \right]^{-1} \mbox{ and} \label{eq:TimeStep_Advection} \\
&\Delta t_\mathrm{diff} = C_{\rmn{CFL}} \left[ \max\mo\left(\frac{D_{\!p\!p}(p)}{p^2}\right) \right]^{-1} \label{eq:TimeStep_Diffusion},
\end{align}
respectively, where $\Delta n$ is the density change of the background gas and
the parameter $C_{\rmn{CFL}}$ is the Courant--Friedrichs--Lewy number for which
we use $C_{\rmn{CFL}}=0.7$ in our simulations.  In principle, the maxima in
equations~\eqref{eq:TimeStep_Advection} and \eqref{eq:TimeStep_Diffusion} have to be
evaluated for all momentum bins, i.e. for $i\in\left[0,\, N{-}1\right]$.  However, in the
absence of Fermi-I \mbox{(re)acceleration}, the momentum range of the advection
and diffusion operator decreases due to rapidly cooling of the spectrum at low
and high momenta. We therefore cut the spectrum at $f_\mathrm{cut}$ below which
we treat numerical values of the spectrum as zero. Hence, there is a low- and a
high-momentum cutoff
\begin{align}
&p_\mathrm{lcut} = \min\mo\left(\left\{p: f(p) \geq f_\mathrm{cut} \right\} \right) \mbox{ and} \label{eq:Momentum_lcut}\\
&p_\mathrm{hcut} = \max\mo\left(\left\{p: f(p) \geq f_\mathrm{cut} \right\} \right), \label{eq:Momentum_hcut}
\end{align}
respectively, and the related indices of the momentum bins
\begin{align}
&i_\mathrm{lcut} = \max\mo\left[0, \min\mo\left(\left\{i: p_i < p_\mathrm{lcut} \right\}\right) - 2\right] \mbox {and} \label{eq:Index_lcut}\\
&i_\mathrm{hcut} = \min\mo\left[N, \max\mo\left(\left\{i: p_i > p_\mathrm{lcut} \right\}\right) + 3\right] \label{eq:Index_hcut}
\end{align}
in between which the maxima in equations~\eqref{eq:TimeStep_Advection}
and \eqref{eq:TimeStep_Diffusion} have to be evaluated, i.e. 
for $i\in\left[i_\mathrm{lcut},\,i_\mathrm{hcut}{-}1\right]$. We consider two extra bins in
equations~\eqref{eq:Index_lcut} and \eqref{eq:Index_hcut} due to the ghost cells of
the advection operator. The cutoff momenta and the related indices are
calculated after every time step.

For clarity, we provide a synopsis of all important momenta and related bin
indices:
\begin{itemize}
\item $p_\mathrm{min}$ and $p_\mathrm{max}$ are the minimum and maximum momenta of our momentum grid, respectively. The corresponding indices are $i_\mathrm{min} = 0$ and $i_\mathrm{max}=N{-}1$.

\item $p_\mathrm{lcut}$ ($p_\mathrm{hcut}$) describes the momentum below (above) which the spectrum is treated as zero. The corresponding indices $i_\mathrm{lcut}$ and $i_\mathrm{hcut}$ account for the ghost cells of the advection operator and are given in equations~\eqref{eq:Index_lcut} and \eqref{eq:Index_hcut}.

\item $p_\mathrm{low}$ and $p_\mathrm{high}$ denote the transition momenta between the numerical and the analytical solution for the low- and high-momentum regime, respectively. The definition is given in section \ref{sec:CombiningAnalyticalNumericalSolution}.

\item $p_\mathrm{cool}$ is the momentum related to inverse Compton and synchrotron cooling in the analytical solution. In the case of freely cooling it coincides with the high-momentum cooling cutoff. In the case of Fermi-I (re)acceleration and injection it is the transition momentum from a source dominated to a steady-state spectrum (see section~\ref{sec:AnalyticSolutionHighMomenta}).

\item $p_\mathrm{acc}$ is the maximum momentum of Fermi-I \mbox{(re)}acceleration where spatial diffusion and cooling balance each other.

\item $p_\mathrm{inj}$ is the injection momentum of Fermi-I \mbox{(re)}acceleration where the non-thermal spectrum is transitions to the non-thermal spectrum.

\end{itemize}

\subsubsection{Modelling Fermi-I (re)acceleration}
We develop an algorithm to account for the Fermi-I process on our tracer
particles and aim at reconstructing the discontinuous Rankine--Hugoniot jump
conditions on the Lagrangian particle trajectories with the aid of a shock
finder in a hydrodynamical scheme.  To this end, we use the adaptive moving-mesh
code \textsc{arepo} \citep{Springel2010} with CR
protons \citep{Pfrommer2017a} and employ the shock finder by \cite{Schaal2015},
which detects cells in the pre-shock region, the shock surface, and the
post-shock zone. The shock direction is determined by the normalized negative
gradient
\begin{align}
\vec{n}_s = -\frac{\bs{\nabla} \tilde{T}}{\abs{\bs{\nabla} \tilde{T}}}
\end{align}
of the pseudo temperature which is given by
\begin{align}
k\tilde{T} = \frac{\mu m_\mathrm{p}(P_\mathrm{th} + P_\mathrm{crp})}{\rho},
\end{align}
where $\mu$ is the mean molecular weight, $m_\mathrm{p}$ is the proton mass,
$\rho$ is the gas mass density, and $P_\mathrm{th}$ and $P_\mathrm{crp}$
denote the thermal and CR proton pressure, respectively. Cells of the {\em
shock zone} are identified by (i) converging flows, i.e. they have a negative
velocity divergence, while (ii) spurious shocks are filtered out and (iii) the
algorithm applies a safeguard in the form of a lower limit to the temperature
and density jump (from pre- to post-shock quantities) to prevent false-positive
detections of numerical noise. The {\em shock surface} cell is identified with
the cell in the shock zone that shows a maximally converging flow along the
shock direction. Pre- and post-shock quantities are obtained from the first
cells outside the shock zone in the direction of shock propagation and opposite
to it, respectively. The algorithm determines the Mach number $\mathcal{M}$ by
the pressure jump and calculates a fraction $\zeta_\mathrm{e}(\theta)$ of
the shock-dissipated energy $E_\mathrm{diss}$ that is converted into the
acceleration of CR electrons,
\begin{align}
\Delta E_\mathrm{cre} = \zeta_\mathrm{e}(\theta) E_\mathrm{diss}. \label{eq:Energy_accelerated_CR_protons}
\end{align}
Here, $\theta$ is the upstream magnetic obliquity, which is the angle between
the direction of shock propagation and the magnetic field.  In this paper, we
assume an acceleration efficiency of $\zeta_\mathrm{e} = 10^{-3}$. This
corresponds to a ratio of accelerated CR electron to proton energies of $\Delta
E_\mathrm{cre} / \Delta E_\mathrm{crp} = 10^{-2}$ for efficient CR proton
acceleration \citep{Pfrommer2017a}. We defer a discussion of the obliquity
dependent acceleration of CR electrons to future studies. We point out that
our description is flexible and can be easily adapted to include new
particle-in-cell simulation results on the shock acceleration of CR
electrons \citep[e.g.][]{Guo2014a,Guo2014b,Park2015}.

As soon as the tracer particle reaches a shock zone cell, we keep the background
density fixed in order to prevent adiabatic heating before encountering the
shock. When the tracer particle transitions from the shock zone to the shock
surface cell, we first calculate the reaccelerated spectrum if there is any
fossil spectrum and secondly, the directly accelerated spectrum.\footnote{We
store only one spectrum in memory per tracer particle. Therefore, we first need
to evaluate the integral in equation~\eqref{eq:Fermi_I_reacceleration_spectrum}
before computing the primary electron spectrum due to diffusive shock
acceleration.} The ambient density of the tracer particles is then set to the
post-shock gas density. In order to model reacceleration and direct
acceleration, we assume continuous injection as a subgrid model and adopt the
source functions\footnote{We use the terminology {\em acceleration} to describe
the production of CR electrons via diffusive shock acceleration and {\em
injection} to describe the generation of secondaries through hadronic
interactions.}
\begin{align}
Q_\mathrm{reac}(p) &= \frac{f_\mathrm{reac}(p)}{\Delta t} \mbox{ and} \label{eq:Fermi_I_reacceleration_source_function} \\
Q_\mathrm{acc}(p) &= \frac{C_\mathrm{acc}}{\Delta t} p^{-\alpha_\mathrm{acc}} \Theta(p - p_\mathrm{inj}),
\label{eq:Fermi_I_acceleration_source_function}
\end{align}
where $\Delta t$ is the the time difference between two MHD time steps. 

As described above, the efficiency of direct Fermi-I acceleration depends
on the total dissipated energy at the shock, which is numerically broadened to a
few cells in finite-volume codes such as \textsc{arepo}. By contrast, Fermi-I
reacceleration only depends on the amplitude of the fossil electron distribution
in the pre-shock region (see equation \ref{eq:Fermi_I_reacceleration_spectrum}),
which is known at the shock surface cell. In both cases, the slope is solely
determined by the Mach number.

To model direct acceleration, we calculate and apply the source function for
acceleration of equation~\eqref{eq:Fermi_I_acceleration_source_function} for
every time step during which the tracer particle resides in a shock surface or
in the post-shock cells for the numerical reasons given above.  By contract, the
source function for reacceleration
(equation~\ref{eq:Fermi_I_reacceleration_source_function}) is only applied
during one MHD time step after the tracer particle has encountered the shock
surface cell.

We calculate $\alpha_\mathrm{acc}$ from the density jump at the
shock, $r~=~\rho_\mathrm{post} / \rho_\mathrm{pre}$, where the pre-shock density
communicated to the shock cell via the \textsc{arepo} shock finder, and the
post-shock density is obtained via
\begin{align}
\rho_\mathrm{post} = \rho_\mathrm{pre} \frac{(\gamma_\mathrm{eff} + 1) \mathcal{M}^2}{(\gamma_\mathrm{eff} - 1) \mathcal{M}^2 + 2}, \label{eq:PostShock_Density}
\end{align}
where the effective adiabatic index is given by
\begin{align}
\gamma_\mathrm{eff} =
\frac{\gamma_\mathrm{crp} P_\mathrm{crp} + \gamma_\mathrm{th} P_\mathrm{th}}{P_\mathrm{crp} + P_\mathrm{th}} \label{eq:Effective_Adiabatic_Index}
\end{align}
with $\gamma_\mathrm{th} = 5/3$ for gas and $\gamma_\mathrm{crp} = 4/3$ for CR protons.

In order to determine the energy of the freshly accelerated CR electrons, we
demand its energy density to be a fixed fraction of the freshly accelerated CR
proton energy density at the shock. In practice, we attach the accelerated
spectrum to the thermal Maxwellian,
\begin{align}
f_\mathrm{th}(p) = 4 \pi n_\mathrm{e, th} \left(\frac{m_\e c^2}{2 \pi
k_\mathrm{B} T}\right)^{3/2} p^2 \exp(-\frac{m_\e c^2 p^2}{2 k_\mathrm{B}
T})
\end{align}
at the injection momentum $p_\mathrm{inj}$ which determines the normalization
\begin{align}
C_\mathrm{acc} = f_\mathrm{th}(p_\mathrm{inj}) p_\mathrm{inj}^{\alpha_\mathrm{acc}}.
\end{align} 	
We use this normalization and the energy of accelerated CR electrons $\Delta E_\mathrm{cre}$, see equation~\eqref{eq:Energy_accelerated_CR_protons}, to determine the injection momentum by the condition
\begin{align}
\int_0^\infty f_\mathrm{th}(p_\mathrm{inj}) p^{\alpha_\mathrm{acc}}_\mathrm{inj} p^{-\alpha_\mathrm{acc}}\Theta(p - p_\mathrm{inj}) E_\mathrm{e, kin}(p) \mathrm{d}p = \frac{\Delta E_\mathrm{cre}}{V_\mathrm{cell}},
\end{align}
where $E_\mathrm{e, kin} (p) = \left[\sqrt{1 + p^2} - 1 \right] m_\e c^2$ is the kinetic energy and $V_\mathrm{cell}$ is the volume of the \textsc{arepo} cell, in which the particle resides.

\subsection{Analytical solutions}
\label{sec:Analytical_Solutions}
\begin{figure}
	\includegraphics[width=\columnwidth]{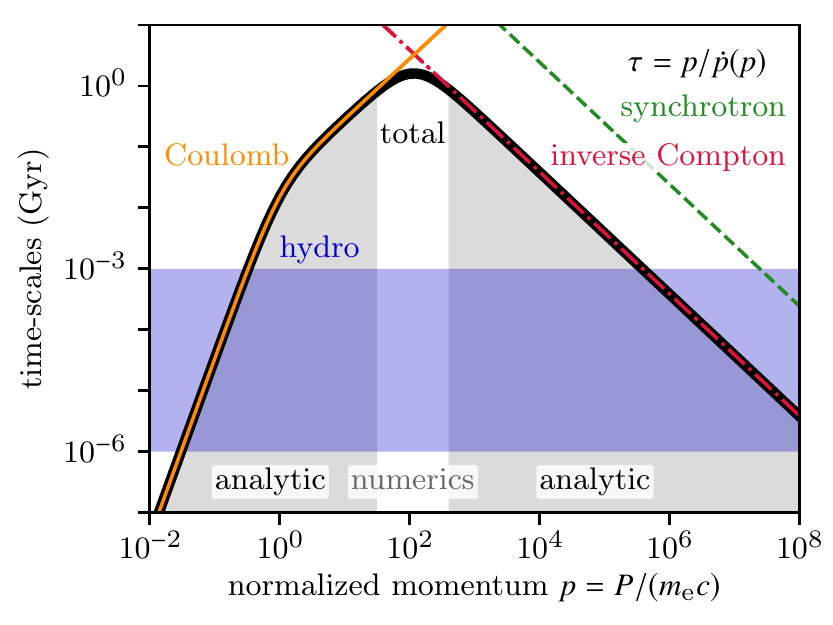} \caption{Characteristic
	time-scales for electron cooling ($n_\mathrm{gas}
	= \SI{e-3}{\per\centi\meter\cubed}$, $B = \SI{5}{\micro\gauss}$,
	$\varepsilon_\rmn{ph} = 6 \varepsilon_\mathrm{cmb}$, $z=0$) and
	typical hydrodynamical time steps adopted in simulations of the ISM and
	galaxy formation (blue band). The grey area shows the ranges where
	either Coulomb or inverse Compton plus synchrotron cooling dominate and
	where analytical solutions can be used. Transition momenta of numerical
	and analytical solutions are $p_\mathrm{low} = \num{3e1}$ and
	$p_\mathrm{high}=\num{4e2}$.}  \label{fig:Characteristic_Timescales}
\end{figure}
The time-scale of all electron cooling processes decreases for low and for high
momenta as can be seen in Figure~\ref{fig:Characteristic_Timescales}, where we show
the cooling times as a function of momentum. Hence, for very low momenta and
very high the cooling time-scales become smaller than the typical
time step of an MHD simulation. In order to have an efficient calculation of the
CR electron spectrum, which advances on time steps similar to the MHD time step, we
use analytical solutions for low and high momenta together with the fully
numerical treatment for intermediate momenta. We call the combination of both
treatments {\em semi-analytical solution}.

\subsubsection{General solutions}
We follow the derivations described by \cite{Sarazin1999} which we summarise
here. The starting point for the analytical solution of the cooling term in
equation~\eqref{eq:CRe_1D_FP_full} is the momentum loss of an individual
electron. Its momentum is shifted from the initial momentum $p_\mathrm{ini}$ to
the momentum $p$ during a time interval of $\Delta t$
\begin{align}
\int_{p_\mathrm{ini}}^p \frac{1}{\dot{p}(p')} \dd{p'} = \Delta t. \label{eq:Cooling_Integral}
\end{align}
Equation~\eqref{eq:Cooling_Integral} is solved for the initial momentum
$p_\mathrm{ini}(p, \Delta t)$, which is used in the analytical solution of the
cooled spectrum
\begin{align}
f(p,t_0 + \Delta t) = f(p_\mathrm{ini}(p,\Delta t), t_0) \frac{\dot{p}(p_\mathrm{ini}(p, \Delta t), t_0 + \Delta t)}{\dot{p}(p, t_0)}. \label{eq:Cooled_Spectrum_Analytic_General}
\end{align}
The cooled spectrum can be interpreted as a momentum shift of the initial
spectrum at time $t_0$ multiplied with a momentum-dependent cooling factor. If
there is no initial spectrum at $t_0$ and if the source function $Q(p, t)$ is
constant and continuous in time, the spectrum after time $t$ is self-similar:
\begin{align}
f_\mathrm{self}(p, t) = f_\mathrm{steady} (p)
- f_\mathrm{steady} (p_\mathrm{ini}(p, t)) \frac{\dot{p}(p_\mathrm{ini}(p, t))}{\dot{p}(p)},
\label{eq:SelfSimilar_Spectrum_General}
\end{align} 
where we use the steady-state solution
\begin{align}
f_\mathrm{steady}(p) = \frac{1}{|\dot{p}(p)|} \int_p^\infty Q(p) \dd{p}. \label{eq:SteadyState_Spectrum_General}
\end{align}
This means that the self-similar solution is derived by subtracting the cooled
steady-state solution from the original steady-state solution. The self-similar
spectrum consists of three characteristic momentum ranges, i.e. low,
intermediate, and high momenta. For low and high momenta, where the cooling
times are smaller than the current time step, the spectrum is already in steady
state. In the intermediate momentum range, the spectrum is dominated by the
source spectrum as we show later.

The analytical solutions of the cooled spectrum in
equation~\eqref{eq:Cooled_Spectrum_Analytic_General} and of the self-similar spectrum
in equation~\eqref{eq:SelfSimilar_Spectrum_General} need a functional representation
of the spectrum at time $t_0$ for the entire momentum range. As the
spectrum is calculated on a discrete momentum grid with piecewise constant
values, we calculate an interpolation function at every time with the Steffen's
method \citep{Steffen1990}, which is cubic and monotonic between neighbouring
discrete momenta.  This interpolation function is used to calculate the analytic
solution after a time step $\Delta t$.

In the following, we present the analytical solutions for both low and high
momenta. We use a source function
$Q(p)= \dot{C}_\mathrm{acc}\,p^{-\alpha_\mathrm{acc}}$ for the self-similar
solution of acceleration and cooling. We note that we have
$\dot{C}_\mathrm{acc}=C_\mathrm{acc}/\Delta t$ for our discretisation and that a
source function for injection by hadronic processes with
$Q(p)=\dot{C}_\mathrm{inj}\,p^{-\alpha_\mathrm{inj}}$ gives similar results for the
self-similar solution. Note that the self-similar solution is not used in our
code (see also section~\ref{sec:AnalyticSolution_SourceCooling}) but in order to
compare simulation results to their analytic solutions.

\subsubsection{Solution for low momenta}
\begin{figure}
	\includegraphics[width=\columnwidth]{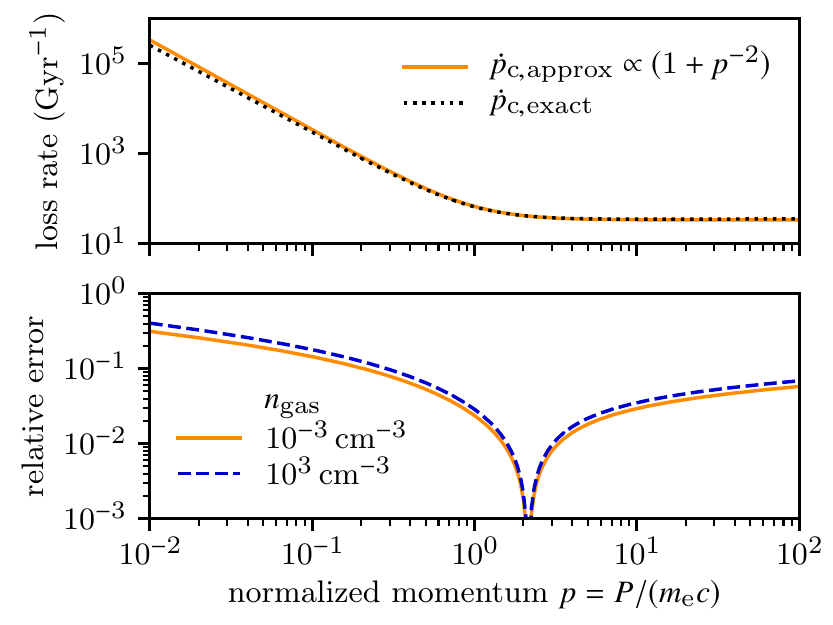}
	\caption{Comparison between exact and approximate formulae for Coulomb losses
		(equations~\eqref{eq:Coulomb_loss_rate_exact}
		and \eqref{eq:Coulomb_loss_rate_approximate}). The top panel shows the
		loss rates for a gas of density $\SI{1e-3}{\per\cubic\cm}$ and the
		bottom panel shows the relative error of the approximate formula for two
		different gas densities.}  \label{fig:CoulombCooling_RateApproximation}
\end{figure}
Coulomb losses are dominating at small momenta. The analytical solution requires
calculating the integral in equation~\eqref{eq:Cooling_Integral} and solving for the
initial momentum $p_\mathrm{ini}(p, \Delta t)$. In general, this cannot be done
in closed analytic form for the exact Coulomb loss rate given in
equation~\eqref{eq:Coulomb_loss_rate_exact}. We therefore use an
approximation \citep{Pinzke2013}
\begin{align}
\dot{p}_\mathrm{c}(p) = b_\mathrm{c} \left(1 + p^{-2}\right) \mbox{ with } b_\mathrm{c} = \frac{- 3 \sigma_\mathrm{T} n_\e c}{2} 
\ln( \frac{m_\e c^2}{\hbar \omega_\mathrm{pl}}), \label{eq:Coulomb_loss_rate_approximate}
\end{align}
which is accurate to $<30\%$ for momenta $10^{-2} \leq p \leq 10^2$ as can be
seen in Figure~\ref{fig:CoulombCooling_RateApproximation}. The integral for the
momentum shift in equation~\eqref{eq:Cooling_Integral} evaluated with the approximate
form of the Coulomb loss rate is
\begin{align}
\int \frac{1}{\dot{p}_\mathrm{c}(p)} \dd{p} =\frac{1}{-b_\mathrm{c}}\left[ p - \arctan(p) \right] \approx \frac{1}{-b_\mathrm{c}}\left( \frac{p^3}{3 + p^2} \right),
\end{align}
where we used a Pad\'e approximation \citep{Brezinski1996} in the last step. The
momentum shift due to Coulomb cooling is then given by
\begin{align}
p_\mathrm{ini}(p, \Delta t) = \frac{1}{3}
\begin{aligned}[t]
& \left[a + \left( a^3 + \frac{9}{2} \sqrt{4a^4 + 81 a^2} + \frac{81 a}{2} \right)^{1/3}\right.\\ 
& \left.~+ a^2 \left( a^3 + \frac{9}{2} \sqrt{4a^4 + 81 a^2} + \frac{81 a}{2} \right)^{-1/3} \right],
\end{aligned}
\label{eq:MomentumShift_LowMomenta}
\end{align}
with $a = p^3 \Big/ \left(3 + p^2\right) - b_\mathrm{c} \Delta t$.  The analytical
solution for the cooled spectrum (see
equation~\eqref{eq:Cooled_Spectrum_Analytic_General}) is given by
\begin{align}
f(p, \Delta t) = f[p_\mathrm{ini}(p, \Delta t), 0] \frac{1 + \left[ p_\mathrm{ini}(p, \Delta t) \right]^{-2} }{1+p^{-2}}.
\end{align}
The self-similar spectrum is given by
\begin{align}
\begin{aligned}
&f_\mathrm{self}(p, \Delta t) \\
&= \frac{\dot{C}_\mathrm{acc}}{(1- \alpha_\mathrm{acc})} 
\begin{aligned}[t]
&\left\{\frac{ p^{-\alpha_\mathrm{acc} + 1}}{\dot{p}_\mathrm{c}^{\phantom{2}}(p)}\right. \\
&\left.\phantom{\Bigg\{} - \frac{ \left[p_\mathrm{ini}(p,\Delta t)\right]^{-\alpha_\mathrm{acc} + 1}}{\dot{p}_\mathrm{c}[p_\mathrm{ini}(p,\Delta t)]}  \frac{1 + \left[ p_\mathrm{ini}(p, \Delta t) \right]^{-2} }{1+p^{-2}}
\right\}\end{aligned}
\end{aligned}
\end{align}
where we use the exact Coulomb loss rate for the first term in the bracket in
order to satisfy $f_\mathrm{self} \rightarrow f_\mathrm{steady}$ for $\Delta
t \rightarrow \infty$.

\subsubsection{Solution for high momenta}
\label{sec:AnalyticSolutionHighMomenta}
For large momenta, inverse Compton and synchrotron cooling are dominating and
both loss rates have the same momentum scaling. We define for convenience
the sum of both as
\begin{align}
\dot{p}_\mathrm{ic+s}(p) = \dot{p}_\mathrm{ic}(p) + \dot{p}_\mathrm{s}(p) = p^2 b_\mathrm{ic+s}
\end{align}
where $b_\mathrm{ic+s} = -4 \sigma_\mathrm{T} \left(B^2/8\pi
+ \varepsilon_\rmn{ph}\right) / (3 m_\e c \beta )$ denotes the momentum
independent factor of both loss rates. The momentum shift during a time interval
$\Delta t$ according to equation~\eqref{eq:Cooling_Integral} is
\begin{align}
p_\mathrm{ini}(p, \Delta t) = \frac{p}{1 - p / p_\mathrm{cool}(\Delta t)} \label{eq:MomentumShift_HighMomenta}
\end{align}
where $p_\mathrm{cool}(\Delta t) = \left(-b_\mathrm{ic+s} \Delta t\right)^{-1}$
is the cooling cutoff of IC and synchrotron losses. In the following, we omit
the explicit time dependence of $p_\mathrm{cool}(\Delta t)$.  The analytical
solution of the cooled spectrum (see
equation~\eqref{eq:Cooled_Spectrum_Analytic_General}) is given by
\begin{align}
&f(p, \Delta t)  = \left\{
	\begin{aligned}
	&f\mo \left(\frac{p}{1 - p / p_\mathrm{cool}},0\right) \left(1 - \frac{p}{p_\mathrm{cool}} \right)^{-2}, &&p<p_\mathrm{cool} \\
	& 0, &&p\geq p_\mathrm{cool}
	\end{aligned}  \right.
\end{align}
and the solution of the self-similar spectrum (see equation~\eqref{eq:SelfSimilar_Spectrum_General}) is
\begin{align}	
f_\mathrm{self}(p, \Delta t)
	 = \frac{\dot{C}_\mathrm{acc}  p^{-(\alpha_\mathrm{acc} + 1) }}{b_\mathrm{ic+s} (1 - \alpha_\mathrm{acc})} \left\{
	\begin{aligned}
		& \left( 1 - \frac{p}{p_\mathrm{cool}} \right)^{\alpha_\mathrm{acc} - 1}, &&p<p_\mathrm{cool} ,\\
		& 1, &&p\geq p_\mathrm{cool}.
	\end{aligned}  \right.
\label{eq:SelfSimilar_HighMomenta}
\end{align}

\subsubsection{Adiabatic changes and cooling}
Pure adiabatic changes due to expansion or compression of the background gas
leave the phase space density of the CR electrons invariant \citep{Ensslin2007}. An
initial spectrum of the form
\begin{align}
f_\mathrm{ini}(p) = C p^{-\alpha} \Theta(p - q)
\end{align}
with normalisation $C$, slope $\alpha$ and low-momentum cutoff $q$ transforms into
\begin{align}
f(p) = C \ratio^{(\alpha + 2)/3} p^{-\alpha} &\Theta\Bigl(p - \ratio^{1/3} q\Bigr)
\end{align}
due to an adiabatic change of the background density from $n_\mathrm{ini}$ to
$n$ and $\ratio=n / n_\mathrm{ini}$ denotes the the ratio of final-to-initial
density. Similar to the analytical description for cooling processes, this
evolution can be interpreted as a shift in momentum space from an initial
momentum $p_\mathrm{ini}$ to momentum $p$ by $p_\mathrm{ini}(p, x) = p \ratio^{-1/3}$
and an overall scaling with the factor $\ratio^{2/3}$
\begin{align}
f(p, \ratio) = \ratio^{2/3} f_\mathrm{ini} \Bigl(p \ratio^{-1/3}\Bigr) .
\end{align}
Our code adopts this equation in combination with the analytical description of
radiation and Coulomb cooling processes.  The evolution of the CR electron
spectrum during small time intervals $\Delta t$ and for small density ratios
$\ratio = n(t+\Delta t) / n(t)$ is described by
\begin{align}
f(p, t + \Delta t) = \ratio^{2/3} f\mo \left[p_\mathrm{ini}\mo \left(\frac{p}{\ratio^{1/3}}, \Delta t\right),  t\right]
\frac{\dot{p}\mo \left[p_\mathrm{ini}\mo \left(p \ratio^{-1/3} , \Delta t\right)\right]}{\dot{p}\mo \left(p \ratio^{-1/3}\right)},
\label{eq:AnalyticSolution_CoolingAdiabtic}
\end{align}
where $p_\mathrm{ini}(p, \Delta t)$ denotes the momentum shift due to cooling as
given in equation~\eqref{eq:MomentumShift_LowMomenta} for low momenta and in
equation~\eqref{eq:MomentumShift_HighMomenta} for high momenta.

\subsubsection{Injection, Fermi-I (re)acceleration and cooling}
\label{sec:AnalyticSolution_SourceCooling}
The analytic solution for the case of cooling and CR electron injection, by
hadronic interactions or by our subgrid model of Fermi-I acceleration and
reacceleration, is in principle given by the self-similar solution in
equation~\eqref{eq:SelfSimilar_Spectrum_General} at time $t$.  However, we cannot use
the self-similar solution because (i) injection{ and (re)acceleration source function} and cooling rates are generally
time-dependent, (ii) we need to evolve the previously existing spectrum, and
(iii) we evolve the spectrum on differential time steps $\Delta t$ from time
$t_n$ to $t_{n+1}$. For large momenta with $p / \dot{p}_\mathrm{ic+s}(p) < \Delta
t$, we use the analytic steady-state solution. For the remaining momentum range,
we use an operator-split method. First, we calculate injection and  Fermi-I (re)acceleration during a half
time step
\begin{align}
f\mo \left(p, t_n + \frac{\Delta t}{2}\right) = f(p, t_n) + \frac{\Delta t}{2} Q(p). \label{eq:Analytic_Injection}
\end{align}
We then calculate the effect of cooling and adiabatic changes on $f(p, t_n +
{\Delta t}/2)$ during a full time step. Finally, we account for injection and (re)acceleration during
another half time step to obtain the spectrum at time $t_{n+1}$,
\begin{align}
f(p, t_{n+1}) =
\begin{aligned}[t]
& f\mo\left[p_\mathrm{ini}\mo \left(\frac{p}{\ratio^{1/3}}, \Delta t\right),  t + \frac{\Delta t}{2}\right]
\frac{\dot{p}\mo \left[p_\mathrm{ini}\mo \left(p \ratio^{-1/3} , \Delta t\right)\right]}{\ratio^{-2/3} \dot{p}\mo \left(p \ratio^{-1/3}\right)}\\
&+ \frac{\Delta t}{2} Q(p).
\end{aligned}
\label{eq:AnalyticSolution_CoolingAdiabticInjection}
\end{align}

\subsubsection{Combining analytical and numerical solutions}
\label{sec:CombiningAnalyticalNumericalSolution}
In general, the momentum loss rate $\dot{p}(p)$ is the sum of all loss processes
which complicates the integral in equation~\eqref{eq:Cooling_Integral} and the
analytical solution for $p_\mathrm{ini}(p, \Delta t)$. As we have seen in the
preceding subsections, analytical solutions are possible for both low momenta
where Coulomb losses are dominating and for high momenta where inverse Compton
and synchrotron losses are dominating.  Our code determines the transition
momenta of the numerical and analytical solutions,
\begin{align}
& p_\mathrm{low} = \max\mo\left(\{ p: \tau_\mathrm{c}(p) < \epsilon \tau_\mathrm{b+ic+s}(p) \ \land\ \tau_\mathrm{c}(p) \leq \tau_\mathrm{hyd}  \}\right) \label{eq:Momentum_low} \mbox{ and}\\
&p_\mathrm{high} = \min\mo\left(\{ p: \tau_\mathrm{ic+s}(p) < \epsilon \tau_\mathrm{b+c}(p) \ \land\  \tau_\mathrm{ic+s}(p) \leq \tau_\mathrm{hyd} \}\right) \label{eq:Momentum_high}
\end{align}
for low and high momenta, respectively. We also take the constraints due to
the hydrodynamical time-scale $\tau_\mathrm{hyd}$ into account. The
characteristic cooling time-scales are $\tau_\mathrm{c} =
p/\dot{p}_\mathrm{c}(p)$ for Coulomb losses, $\tau_\mathrm{b} =
p/\dot{p}_\mathrm{b}(p)$ for bremsstrahlung, and $\tau_\mathrm{ic+s} =
p/\dot{p}_\mathrm{ic+s}(p)$ for IC and synchrotron cooling. The transition
momentum is determined by a free parameter, which we set to $\epsilon=0.1$. The
characteristic cooling time-scales and the transition momenta are displayed in
Figure~\ref{fig:Characteristic_Timescales}.

We determine corresponding indices of the transition momentum bins as
\begin{align}
&i_\mathrm{low} = \max\mo\left[0, \max\mo\left(\{i : p_i < p_\mathrm{low} \}\right) - 2\right] \mbox{ and}\label{eq:Index_low} \\
&i_\mathrm{high} = \min\mo\left[N, \min\mo\left(\{i : p_i > p_\mathrm{high} \}\right) + 3\right]. \label{eq:Index_high}
\end{align}
between which the numerical solution is applied, i.e. for the momentum bins $p_i$ with $i\in\left[i_\mathrm{low},\,i_\mathrm{high} \right]$.
Analytical solutions are calculated for low-momentum bins $p_i$ with $i\in\left[0,\,i_\mathrm{low}{+}2\right]$ and high-momentum bins with $i\in\left[i_\mathrm{high}{-}3,\,N{-}1\right]$. At the indices $i_\mathrm{low}{+}2$ and $i_\mathrm{high}{-}3$, we calculate the ratio of numerical to analytical solution in the low- and high-momentum regime
\begin{align}
&C_\mathrm{low} = \frac{\mathcal{A}_\mathrm{adv}^\mathrm{num}(\Delta t) f(p_{i_\mathrm{low} + 2}, t)}{ \mathcal{A}_\mathrm{adv}^\mathrm{ana}(\Delta t) f(p_{i_\mathrm{low} + 2}, t)} \mbox{ and} \\
&C_\mathrm{high} = \frac{\mathcal{A}_\mathrm{adv}^\mathrm{num}(\Delta t) f(p_{i_\mathrm{high} - 3}, t)}{ \mathcal{A}_\mathrm{adv}^\mathrm{ana}(\Delta t) f(p_{i_\mathrm{high} - 3}, t)},
\end{align}
respectively, where $\mathcal{A}_\mathrm{adv}^\mathrm{num}$ is the numerical advection operator and $\mathcal{A}_\mathrm{adv}^\mathrm{ana}$ the analytical advection operator for low and high momenta.

The analytical solutions in the low- and high-momentum regime are multiplied
with these ratios in order to guarantee a continuous spectrum. Hence, the
evolved spectrum at momentum bin $p_i$ is given by
\begin{align}
f(p_i, t+\Delta t) = \left\{
\begin{aligned}
&C_\mathrm{low} \mathcal{A}_\mathrm{adv}^\mathrm{ana}(\Delta t) f(p_i, t) &&\mbox{for } i\in\left[0,\, i_\mathrm{low}{+} 1\right] \\
&\mathcal{A}_\mathrm{adv}^\mathrm{num}(\Delta t) f(p_i, t) &&\mbox{for } i\in\left[i_\mathrm{low}{+}2,\, i_\mathrm{high}{-}3\right] \\
&C_\mathrm{high} \mathcal{A}_\mathrm{adv}^\mathrm{ana}(\Delta t) f(p_i, t) &&\mbox{for } i\in\left[i_\mathrm{high}{-}2,\, N{-}1\right]. \\
\end{aligned}\right.
\end{align}

\section{Idealised one-zone tests}
\label{sec:Idealized_One_Zone_Tests}
In order to demonstrate the validity of \textsc{crest}, we first conduct idealised
one-zone tests. These setups evolve the CR electron spectrum without an MHD simulation,
hence necessary parameters for the spectral evolution are defined by hand. These
tests demonstrate that our code is able to accurately and correctly simulate
adiabatic processes, non-adiabatic cooling, acceleration and diffusion in momentum
space.

\subsection{Adiabatic changes}

\begin{figure}
	\includegraphics[width=\columnwidth]{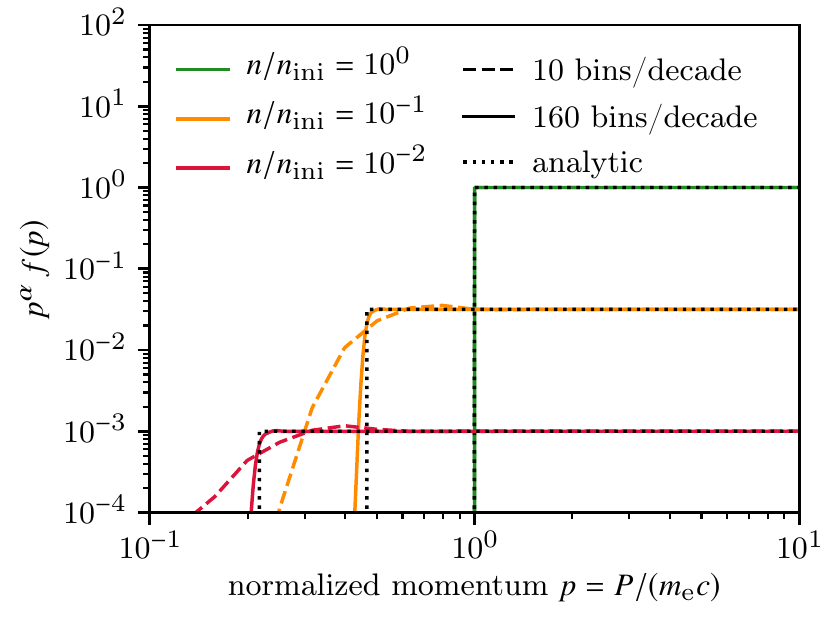} \includegraphics[width=\columnwidth]{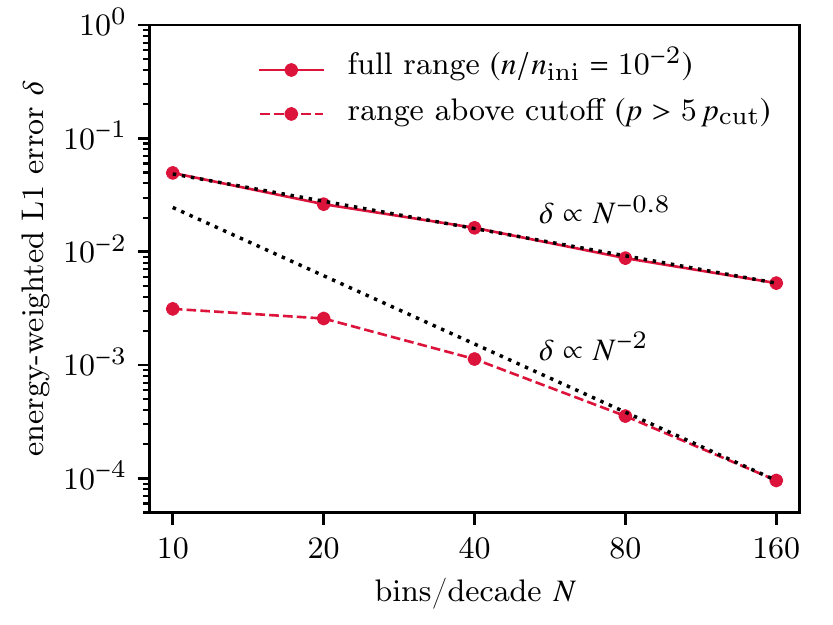} \caption{Adiabatic
	expansion of an initial power-law spectrum with $\alpha = {2.5}$. Top:
	coloured dashed and solid lines represent the simulations with 10 and 160
	bins per decade, respectively. The analytical solutions are shown as
	black dotted lines. Bottom: the energy-weighted relative L1 error for
	the entire momentum range and for momenta much larger than the cutoff of
	the analytical solution $p_\mathrm{cut} = 10^{-2/3}$.}  \label{fig:Adiabatic_Expansion}
\end{figure}
Adiabatic changes are mediated through the velocity divergence terms in
equation~\eqref{eq:CRe_1D_FP_full}. Due to phase space conservation upon adiabatic
changes, a decreasing (increasing) gas density leads to decreasing (increasing)
normalisation and a shift of the CR electron spectrum towards smaller (larger)
momenta. In Figure~\ref{fig:Adiabatic_Expansion}, we follow the evolution of the
spectrum during an adiabatic expansion over an expansion factor of
$10^{-2}$. The energy-weighted L1 error between the simulated spectrum
$f_\mathrm{sim}$ and the analytical spectrum $f_\mathrm{ana}$ is calculated
according to the formula
\begin{equation}
\delta = \frac{\int \left|f_\mathrm{sim}(p) - f_\mathrm{ana}(p)\right| T(p) \dd{p}}{\int f_\mathrm{ana}(p) T(p) \dd{p}}, \label{eq:NumericalError}
\end{equation}
and decreases for increasing number of momentum bins $N$. The error scaling for
the entire momentum range shows the effect of the slope limiter, which uses a
second order accurate scheme for smooth parts of the spectrum and resorts to a
first order scheme near jumps or strong gradients to prevent numerical
oscillations. However, in the range above the cutoff the error scales as
$\delta \propto N^{-2}$ as expected for a second-order accurate numerical
scheme. We note that cooling and momentum diffusion normally lead to a smooth
spectrum without sharp features. Hence, adiabatic changes are calculated with
second-order accuracy.

\subsection{Freely cooling spectrum}
\begin{figure}
\includegraphics[width=\columnwidth]{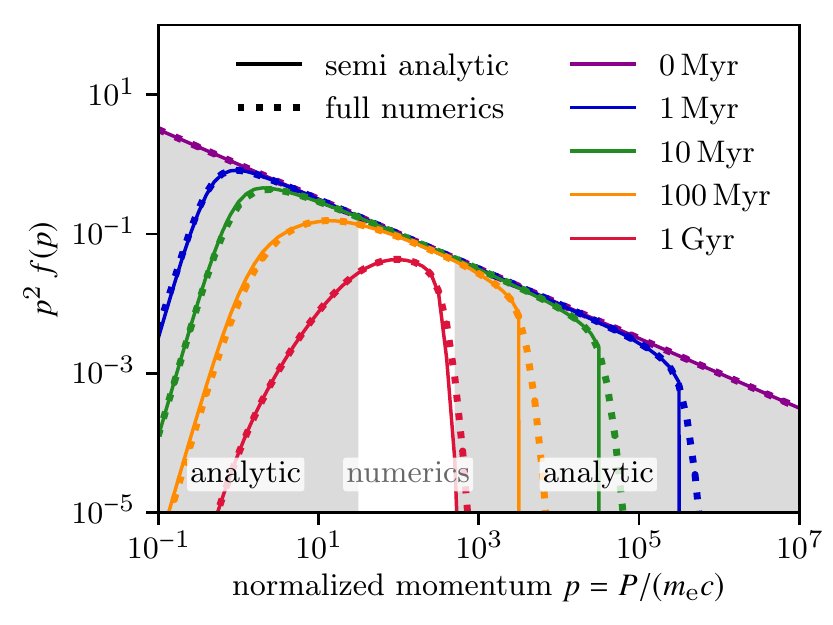}
\caption{Freely cooling power-law spectrum with $\alpha=2.5$. We compare the
	fully numerical and semi-analytical solutions, for which we adopt
	analytical solutions in the shaded momentum range. The simulations use
	10 bins per decade and the relevant parameters are ${n_\mathrm{gas} =
	10^{-3}\,\mathrm{cm}^{-3}}$, $B =\SI{5}{\micro\gauss}$ and
	$\varepsilon_\rmn{ph} =
	6\,\varepsilon_\mathrm{cmb}$.}  \label{fig:OneZone_FreelyCooling}
\end{figure}
\begin{figure}
\includegraphics[width=\columnwidth]{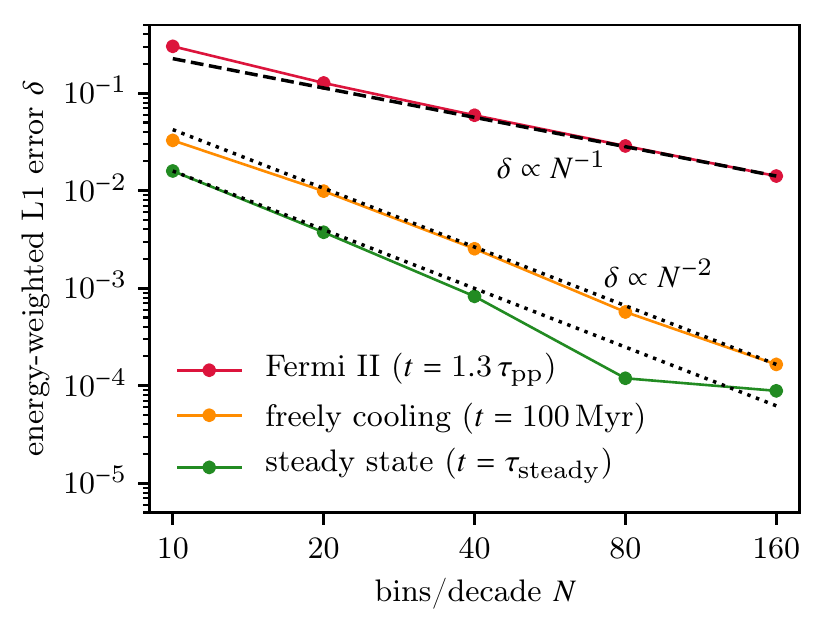}
\caption{Energy-weighted relative L1 errors for cooling, steady-state and Fermi-II
	reacceleration tests. }
\label{fig:OneZone_Error}
\end{figure}
A CR electron spectrum may experience cooling due to Coulomb, bremsstrahlung, inverse
Compton and synchrotron losses. Figure~\ref{fig:OneZone_FreelyCooling} shows the
cooling of an initial power-law spectrum with spectral index of $\alpha = 2.5$
for a setup with 10 bins per decade. We compare the fully numerical solution to
the semi-analytical solution, which uses the analytical solution in the shaded
momentum ranges and the fully numerical solution in the range in between, where
all cooling processes modify the initial power law.  The fully numerical
solution matches the semi-analytical solution except for the high-momentum
cutoff which displays a larger diffusivity for the fully numerical scheme. The
error of the fully numerical solution with $N$ bins is calculated according to
equation~\eqref{eq:NumericalError} where we take the simulation with double
resolution as $f_\mathrm{ana} \approx f_{2N}$. The error scaling is shown in
Figure~\ref{fig:OneZone_Error} and is second-order accurate, i.e. $\delta \propto
N^{-2}$.

\subsection{Steady-state spectrum}
\begin{figure}
\includegraphics[width=\columnwidth]{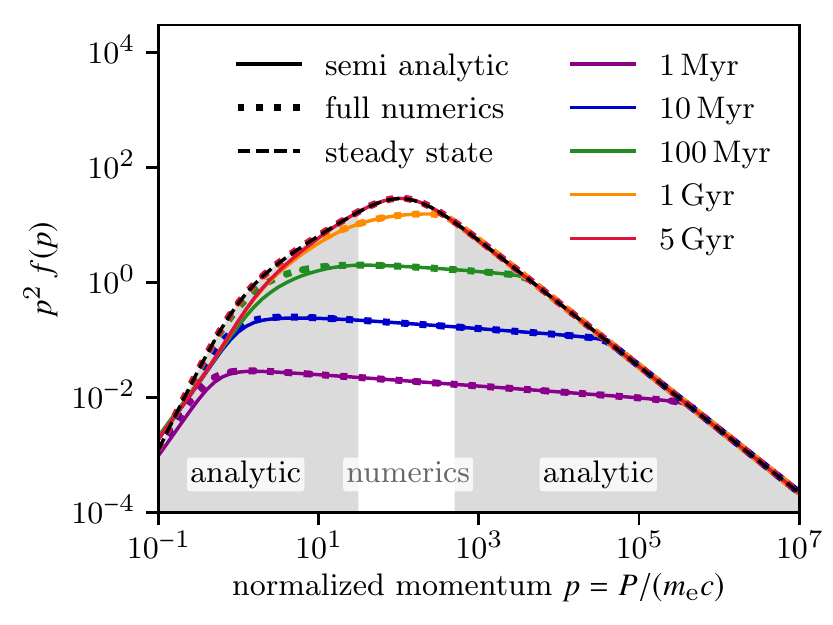}
\caption{Build up of a steady-state spectrum due to continuous injection and
        cooling. The solid line and the dotted lines show the semi-analytical
         and the fully numerical simulations, respectively. The analytical
         steady-state solution is shown with a dashed line. The simulations use
         10 bins per decade and a power-law source function with $\alpha
         = 2.1$. The relevant parameters are $n_\mathrm{gas}
         = \SI{e-3}{\per\cubic\cm}$, $B = \SI{5}{\micro\gauss}$ and
         $\varepsilon_\rmn{ph} =
         6\,\varepsilon_\mathrm{cmb}$.}  \label{fig:OneZone_SteadyState}
\end{figure}
The combination of cooling and continuous source function $Q(p,t)$, e.g. acceleration or injection, in
equation~\eqref{eq:CRe_1D_FP_full} leads to the build up of a self-similar
spectrum. The self-similar spectrum agrees with the steady-state spectrum for
momenta that have smaller cooling time-scales in comparison to the simulation
time. Hence, the self-similar spectrum completely approaches the steady-state
spectrum for very long times. We show this evolution in
Figure~\ref{fig:OneZone_SteadyState} where we compare the results of the fully
numerical and the semi-analytical simulations as well. Both simulations agree
relatively well and approach the steady-state solution. However, there is a
small deviation of the semi-analytical simulation visible in the Coulomb regime
at around $p=1$. This is a consequence of the approximations adopted that enable
an analytical solution for Coulomb cooling. Nevertheless, we prefer the
semi-analytical simulation as it generally outperforms in efficiency in
comparison to the fully numerical simulation (it is faster by a factor of $\sim
10^4$ for this specific setup). The error of the fully numerical solution 
compared to the analytical steady-state solution (see equations 
\eqref{eq:SteadyState_Spectrum_General} and \eqref{eq:NumericalError}) is shown
in Figure~\ref{fig:OneZone_Error} and scales with $\delta \propto N^{-2}$.

\subsection{Fermi-II reacceleration}
\begin{figure}
\includegraphics[width=\columnwidth]{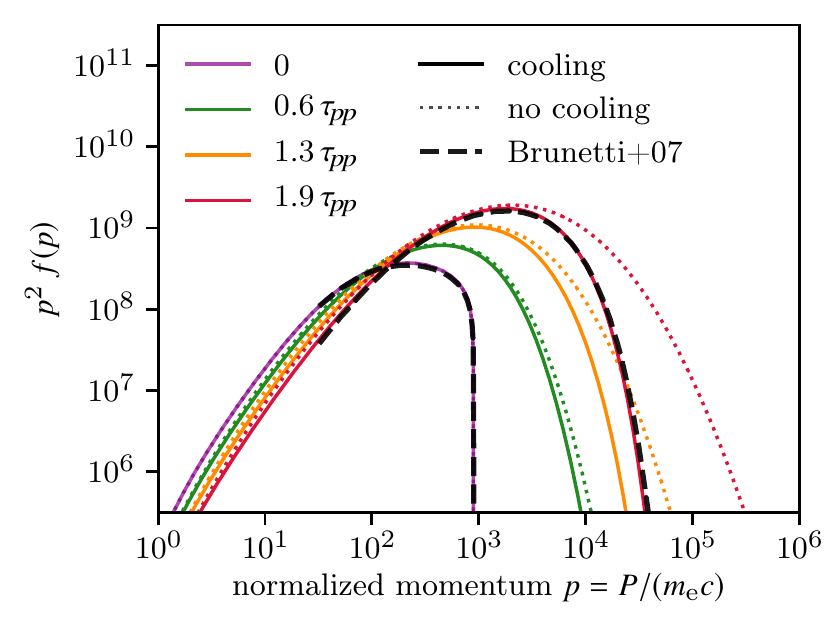}
\caption{Fermi-II reacceleration with and without cooling of a relic spectrum
        in comparison to a reference study \citep{Brunetti2007}. The
	semi-analytical and fully numerical simulations that include cooling are
	indistinguishable. The relevant parameters are $n_\mathrm{gas}
	= \SI{e-3}{\per\cubic\cm}$, $B=\SI{1}{\micro\gauss}$, $\varepsilon_\rmn{ph}
	= \varepsilon_\mathrm{cmb}$ and $\tau_{\!p\!p} = \SI{0.2}{Gyr}$. The
	time steps given in multiples of $\tau_{\!p\!p}$
	are \SIlist{0;127;254;381}{Myr}.}  \label{fig:OneZone_Fermi2_plusCooling}
\end{figure}
In addition to adiabatic changes, cooling, Fermi-I \mbox{(re)}acceleration, and
injection, the CR electron spectrum may experience Fermi-II reacceleration,
which is described by the momentum diffusion terms in
equation~\eqref{eq:CRe_1D_FP_full} and which increases the energy of the spectrum. We
adopt a typical value for the diffusion time of $\tau_{\!p\!p} = \SI{0.2}{Gyr}$
in our tests. In Figure~\ref{fig:OneZone_Fermi2_plusCooling}, we show two
simulations with and without cooling for a high resolution of 160 bins per
decade for Fermi-II reacceleration. Both simulations start with the same initial
spectrum, which we have taken from a study on Fermi-II reacceleration of CR
electrons by \cite{Brunetti2007}. The simulation with cooling approaches a limit
for high momenta where cooling dominates over the reacceleration by the Fermi-II
process. The result of the simulation with cooling matches the reference
simulation by \cite{Brunetti2007} very well. The simulation without cooling
shows the main effect of Fermi-II reacceleration, i.e. diffusion in momentum
space and a shift towards higher particle energies.

Figure~\ref{fig:OneZone_Error} shows the error scaling of the different
simulations with number of bins per momentum decade. The error is calculated
with equation~\eqref{eq:NumericalError}. However, for the simulations of freely
cooling and momentum diffusion, we compare the result at given resolution $f_N$
to the double resolution, i.e. $f_\mathrm{ana} \approx f_{2N}$.  The
implemented Crank--Nicolson scheme is only accurate to first order in momentum
space as can be seen by the $\delta \propto N^{-1}$ scaling. We consider this result
for the Fermi-II reacceleration as a proof of concept. The improvement of the
diffusion operator is straightforward but beyond the scope of this paper.
The simulation of freely cooling and steady state show an error scaling of
$\delta \propto N^{-2}$ which reflects our second-order accurate scheme for
advection with a slope limiter.

\begin{figure*}
\includegraphics[width=\textwidth]{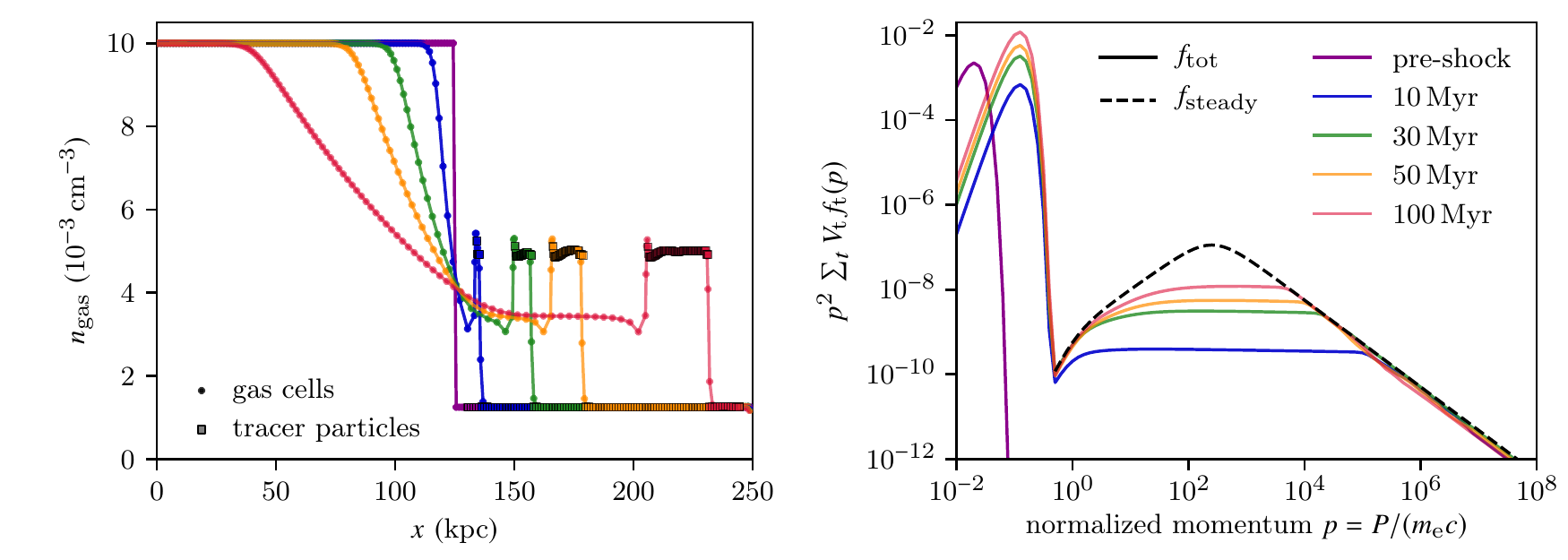}
\caption{1D shock-tube test of a strong shock ($\mathcal{M}=8.43, \alpha_\mathrm{acc} = 2.0$)
        with 200 cells and 100 tracer particles. The left-hand panel shows the
	gas density profiles together with the tracer particles on which we
	evolve the CR electron spectra at different times. The right-hand panel
	shows the different total volume integrated thermal and CR electron
	spectra and the theoretically expected steady-state spectrum (dashed),
	for which we adopt $\alpha_\mathrm{acc}=2.02$ instead of the theoretical
	value $2.0$ to account for the numerical scatter of the shock
	compression ratio (see Figure~\ref{fig:ShockTube1D_ShockStatistics}). We
	adopt the parameters $B = \SI{1}{\micro\gauss}$ and $\varepsilon_\rmn{ph} =
	6\,\varepsilon_\mathrm{cmb}$.}  \label{fig:ShockTube1D_StrongShock_Arepo_CRe_Injection_Cooling}
\end{figure*}

\section{Hydrodynamical simulations}
\label{sec:Simulations_with_arepo}
In addition to idealised one-zone tests, we demonstrate that \textsc{crest}
works in tandem with a hydrodynamical code. To this end, we use the
second-order accurate, adaptive moving-mesh
code \textsc{arepo} \citep{Springel2010, Pakmor2016} for simulations with ideal
MHD \citep{Pakmor2013}. CR protons are modelled as a relativistic fluid with a
constant adiabatic index of $\gamma_\mathrm{crp} = \ 4/3$ in a two-fluid
approximation \citep{Pfrommer2017a}.  We include Lagrangian tracer particles,
which are velocity field tracers \citep{Genel2013} and are passively advected
with the gas and on which we solve the CR electron transport equation in post
processing on every MHD time step.

To assess the validity of our setup, we investigate two different hydrodynamical
scenarios, shock-tube simulations and 3D Sedov--Taylor blast-wave
simulations. This enables us to probe Fermi-I acceleration and reacceleration,
cooling and adiabatic processes in more realistic setups.  The CR electron
spectrum is calculated in post-processing separately for every tracer particle
and the relevant parameters for the spectral evolution are taken from the gas
cells which contain the tracer particles.

\subsection{Shock tubes}
\begin{figure*}
\includegraphics[width=\textwidth]{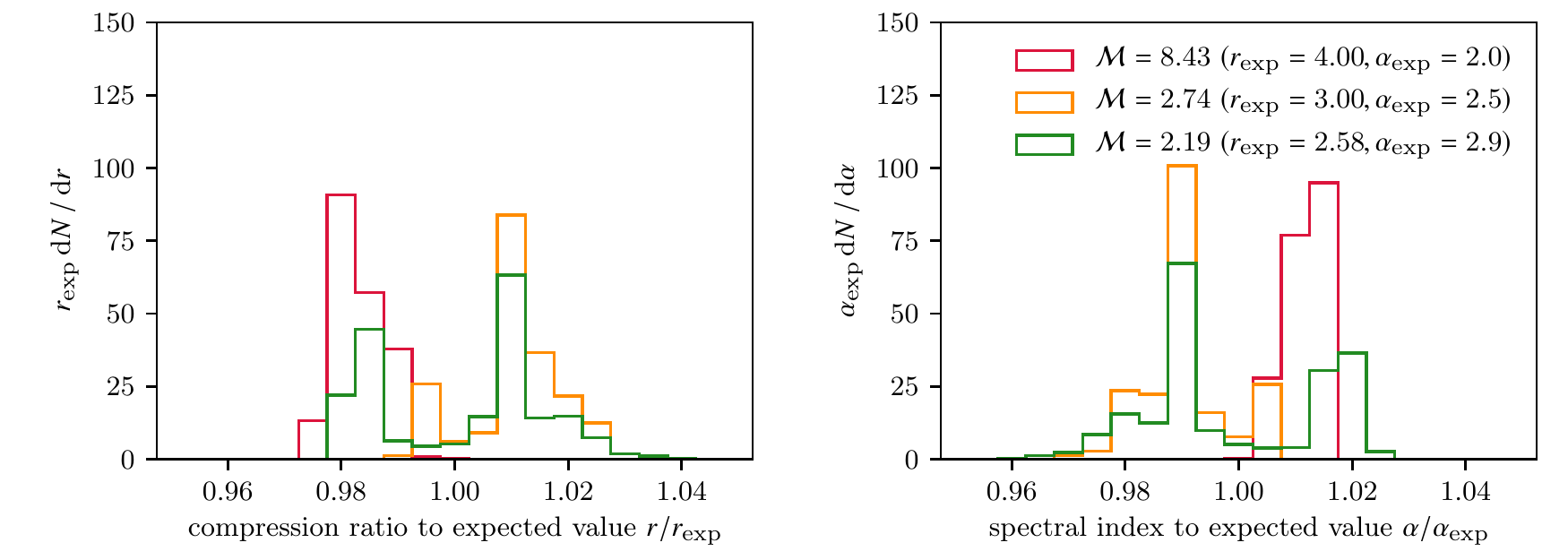}
\caption{Histograms of the compression ratio (left-hand panel) and spectral index
        (right-hand panel), which are both normalised to their expected values
	for three different 1D shock-tube tests, which use the parameters given
	in table~\ref{tab:ShockTube_InitialValues}. The histograms account for
	all tracer particles at all time steps provided they experience
	an acceleration event.}  \label{fig:ShockTube1D_ShockStatistics}
\end{figure*}

\begin{figure*}
\includegraphics[width=\textwidth]{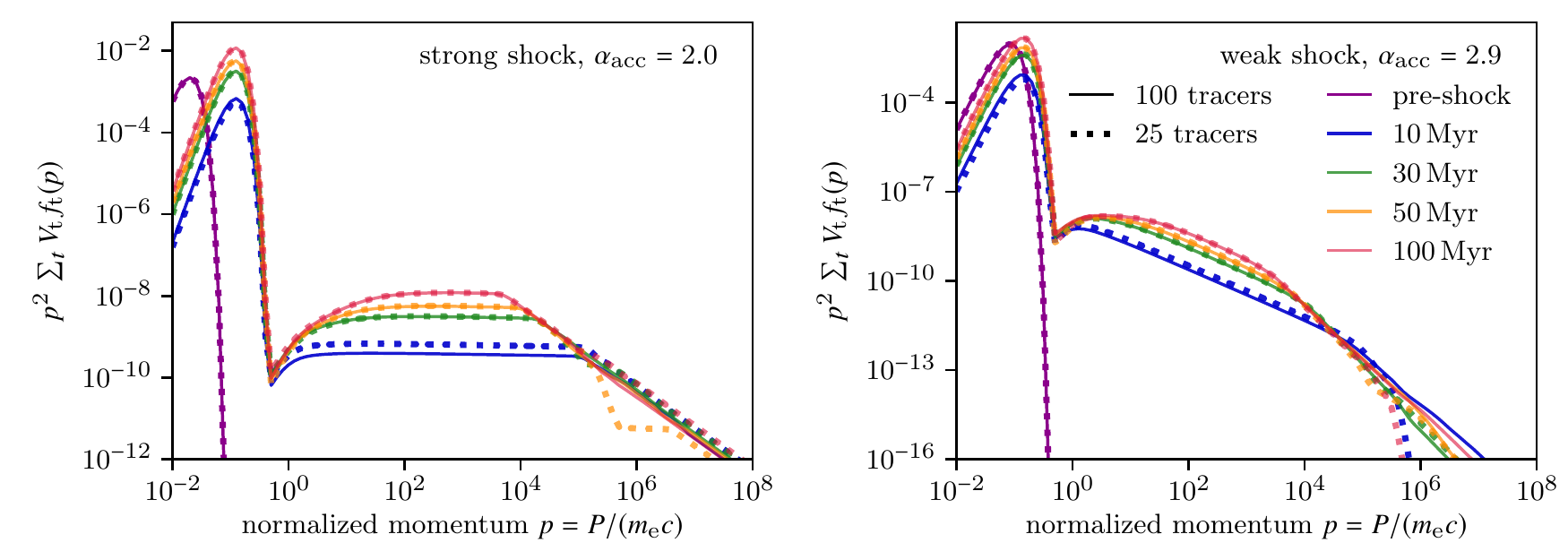}
\caption{Tracer particle resolution study for 1D shock tubes of a strong shock
	(left-hand panel) and a weak shock (right-hand panel). The solid lines
	display the simulation with 100 tracer particles, which is for the
	strong shock identical to the right-hand panel in
	Figure~\ref{fig:ShockTube1D_StrongShock_Arepo_CRe_Injection_Cooling}, and
	the dotted line displays the simulation with 25 tracer
	particles. Low-resolution runs show temporary dips, but generally match
	the high-resolution runs
	well.}  \label{fig:ShockTube1D_TracerResolutionComparison}
\end{figure*}
\begin{figure*}
\includegraphics[width=\textwidth]{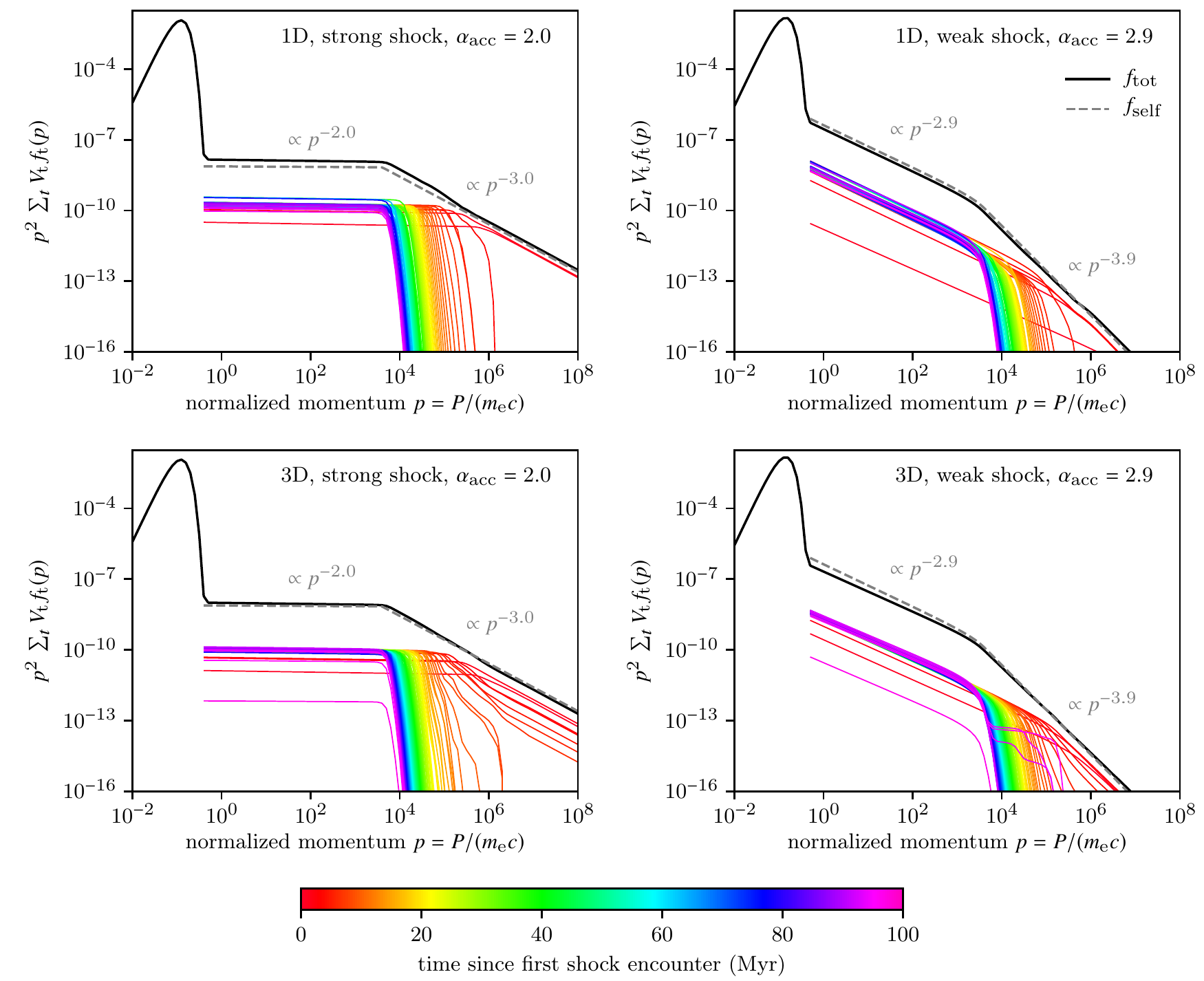}
\caption{Total thermal and CR electron spectra at \SI{100}{Myr} (black solid lines) and partial 
	spectra of 100 time intervals since first shock encounter (coloured thin
	solid lines) for 1D and 3D shock-tube simulations (top and bottom,
	respectively) of strong and weak shocks (left- and right-hand panels,
	respectively). The theoretically expected steady-state spectrum (dashed)
	matches the total spectrum very well. Inverse Compton and synchrotron
	cooling lead to steeper spectra for large momenta. Note that Coulomb and
	bremsstrahlung cooling is neglected here.}
\label{fig:ShockTube_1D_3D_InjectionCooling_Comparison}
\end{figure*}

\begin{figure*}
\includegraphics[width=\textwidth]{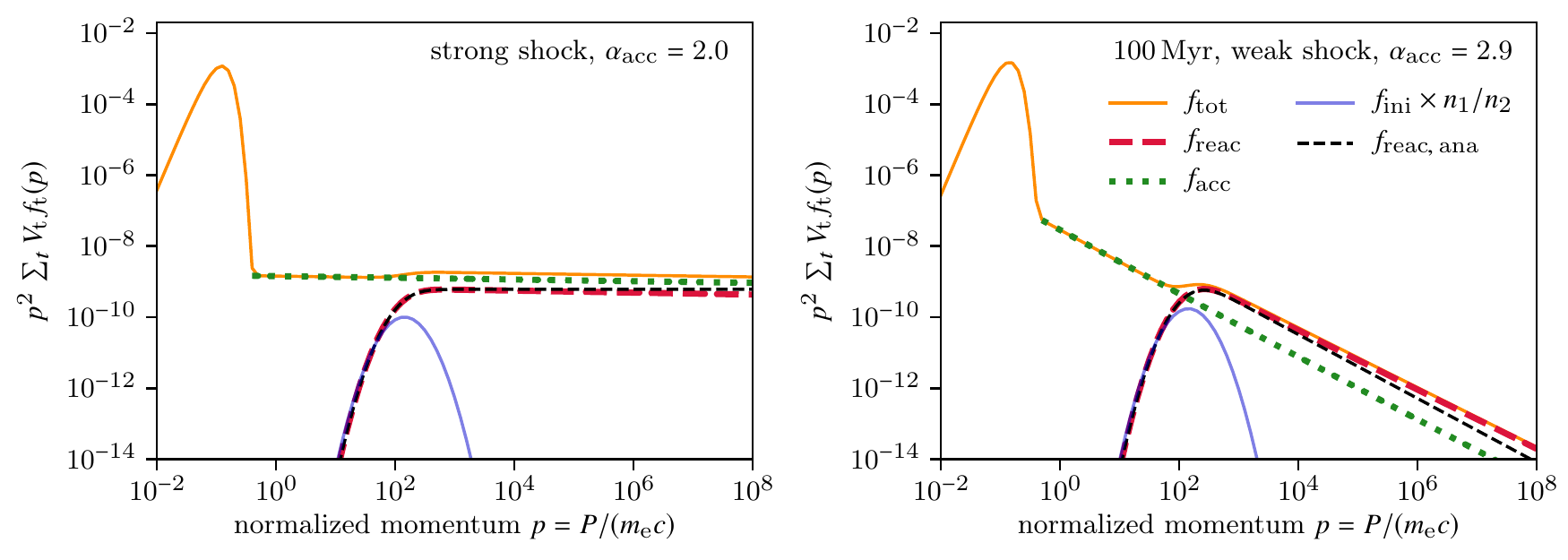}
\caption{Comparison of direct acceleration (green dotted) and reacceleration
        (red dashed) without cooling in 1D shock-tube simulations for a strong
	and a weak shock (left- and right-hand panels, respectively). We show
	the initial relic spectrum (blue) and the total spectrum after the
	acceleration event (orange). The theoretically expected reacceleration
	spectrum matches the simulation very well.}
\label{fig:ShockTube1D_Reacceleration_Comparison}
\end{figure*}

\begin{table}
	\caption{Initial values of our shock-tube setups. The parameters
	$n_\mathrm{L} = \SI{1e-2}{\per\cubic\cm}$, $P_{\mathrm{th, L}}
	= \SI{2.62e-11}{\erg\per\cm\cubed}$ and $P_{\mathrm{crp, L}} /
	P_{\mathrm{th, L}}=2$ for the left initial state and $n_\rmn{R}
	= \SI{0.125e-2}{\per\cubic\cm}$ and $P_{\mathrm{crp, R}} / P_{\mathrm{th,
	R}}=1$ for the right initial state are the same for all simulations.}
	\begin{tabular}{lllll}
		\hline
		$P_\mathrm{th, R}\ (\si{\erg\per\cm\cubed})$ & $P_\mathrm{th, L} / P_\mathrm{th, R}$ & $\mathcal{M}$ & $r$ & $\alpha_\mathrm{acc}$ \\
		\hline\noalign{\smallskip}
		\num{1.06e-13} & 247.0 & \num{8.43} & \num{4.0} & \num{2.0} \\
		\num{3.14e-13} & ~23.4 & \num{2.74} & \num{3.0} & \num{2.5} \\
		\num{1.89e-12} & ~13.9 & \num{2.19} & \num{2.58} & \num{2.9}\\
		\hline
	\end{tabular}
	\label{tab:ShockTube_InitialValues}
\end{table}
First, we perform a series of shock-tube tests \citep{Sod1978} in \textsc{arepo}
with various shock strengths. The fluid is composed of gas and CR protons and we
take CR acceleration at the shock in account \citep{Pfrommer2017a} with CR proton shock
acceleration efficiency of $\zeta_\mathrm{crp} = 0.1$.  In our 1D setups, we use
a box with \SI{250}{\kilo\parsec} side length and 200 cells. In addition 100
tracer particles are located in the initial state on the right-hand side. For
the 3D simulations, we use a box of dimension $250 \times 25 \times
25\,\si{\kilo\parsec}$ with $200 \times 20 \times 20$ cells and $100 \times
10 \times 10$ tracer particles in the initial state on the right-hand side. The tracer particles initially only contain a thermal electron spectrum. The
initial states of the Sod shock-tube problem are laid down in
table~\ref{tab:ShockTube_InitialValues}. We vary the thermal pressure
$P_{\mathrm{th, R}}$ in order to obtain a desired Mach number $\mathcal{M}$ and
1D acceleration spectral index $\alpha_\mathrm{acc}$, which is a function of the
shock compression ratio $r$, i.e. $\alpha_\mathrm{acc} = (r + 2)/(r-1)$.

Figure~\ref{fig:ShockTube1D_StrongShock_Arepo_CRe_Injection_Cooling} shows a 1D
shock-tube test of a strong shock
($\mathcal{M}=8.43, \alpha_\mathrm{acc}=2.0$). The left-hand panel shows the gas
density together with the tracer particles for different snapshots. The
right-hand panel shows the thermal and CR electron spectra as a volume
integrated sum of the tracer particle spectra, which have thermal spectra in the
initial state. Except for the initial state at $t=0$, we sum up only spectra
from those particles that have already encountered the shock front.  Due to the
initial inhomogeneity, a shock develops and propagates into the state on the
right-hand side where the first tracer particle crosses the shock after $\sim
5\,\mathrm{Myr}$. As soon as a tracer particle encounters the shock front, CR
electron acceleration is triggered, i.e. we use a source term of the form
$Q_\mathrm{e}(p) \propto p^{-\alpha_\mathrm{acc}}$ in the transport equation
(see equations~\eqref{eq:CRe_1D_FP_full} and \eqref{eq:Fermi_I_acceleration_spectrum}).
The CR electron spectra experience losses due to Coulomb, bremsstrahlung,
inverse Compton, and synchrotron interactions at the same time. Hence, the total
spectrum has the form of a self-similar spectrum
(see equation~\eqref{eq:SelfSimilar_Spectrum_General}).

The spectrum in
Figure~\ref{fig:ShockTube1D_StrongShock_Arepo_CRe_Injection_Cooling} approaches a
steady state in the momentum regime, which has a shorter cooling time in
comparison to the time since the first shock encounter. The total spectrum is
similar to our idealised one-zone test, which simulates only one spectrum that
experiences continuous cooling and injection. However, the simulation
with \textsc{arepo} uses many tracer particles which experience acceleration only
for limited amount of time when the particle resides in a shock surface or post
shock cell of the hydrodynamical simulation. This clearly demonstrates that the
combination of numerical and analytical solutions produces an effective, stable
and accurate algorithm.

As pointed out before, the spectral index $\alpha_\mathrm{acc}$ of the
accelerated spectrum depends on the shock compression ratio which is subject to
numerical inaccuracies. In Figure~\ref{fig:ShockTube1D_ShockStatistics}, we show
histograms for ratios of the numerically obtained value of shock compression to
its expected value $r / r_\mathrm{exp}$ and ratios of the numerically obtained
value of spectral index to its expected value $\alpha / \alpha_\mathrm{exp}$ for
three different shock strengths (or equivalently Mach numbers). Here, we
calculate the shock compression ratio with equation~\eqref{eq:PostShock_Density},
which depends on the Mach number and which is formally only accurate for a
single polytropic fluid. However, this calculation yields better results in
comparison to the shock compression ratio directly calculated by
the \textsc{arepo} shock finder. The resulting numerical error for the Mach
number is typically better than one per cent (and deteriorates up to two per cent for weak
shocks).

A resolution test of the number of tracer particles is shown in
Figure~\ref{fig:ShockTube1D_TracerResolutionComparison}, which displays the
total spectra for 25 and 100 tracer particles for strong and weak shocks. The
low-resolution spectra can show temporary dips due to poor sampling of the
tracer particles in space, in particular at high momenta. However,
low-resolution runs are stable and reproduce the general result of
high-resolution runs. This demonstrates that our code produces stable and
accurate results (only limited by the sampling rate) with respect to a coarser
sampling of the tracer particles than the gas cells.

The total spectrum is a sum of all tracer particle spectra as we show in
Figure~\ref{fig:ShockTube_1D_3D_InjectionCooling_Comparison}. There, we plot the
results of 1D and 3D simulations for strong and weak shocks. Note that we only
consider inverse Compton and synchrotron cooling for clarity. Each panel shows
the total spectrum, the theoretically expected self-similar spectrum, and
partial sums of spectra of 100 equally spaced time intervals since the first
shock encounter. Those particles that have most recently crossed the shock (red
lines) experience simultaneously acceleration and cooling and show a
self-similar spectrum. The spectra of those particles that have encountered the
shock some time ago (orange to purple lines) show an exponential high-momentum
cutoff resulting from the freely cooling CR electron population. The total
spectrum has the slope of the acceleration spectrum for those momenta which have
cooling times longer than \SI{100}{Myr}, i.e. $p \lesssim 10^4$ for these
setups. At larger momenta, $p \gtrsim 10^4$ , the slope of the total spectrum
steepens to {$\alpha_\mathrm{acc} + 1$} as expected from
equation~\eqref{eq:SelfSimilar_HighMomenta}. The total spectra for all setups
match the theoretically expected self-similar spectra very well, although slight
deviations are visible. These follow from the numerical scatter of the shock
compression ratio in \textsc{arepo}. We note that the computation of the CR
electron spectrum with \textsc{crest} is faster than the hydrodynamical
simulation by a factor of about 20 in the 3D shock-tube simulations.

In addition to direct acceleration of primary CR electrons at the shock, a
previously existing non-thermal CR electron population can be
reaccelerated at the shock. We show the resulting spectra for a strong and a
weak shock after \SI{100}{Myr} in
Figure~\ref{fig:ShockTube1D_Reacceleration_Comparison}. The setups are similar to
the simulations presented above except for the previously existing non-thermal
relic spectrum and except for the fact that we deactivated CR electron cooling
for clarity here. As soon as a tracer particle encounters the shock, it
experiences both, reacceleration of the initial relic spectrum
and direct acceleration of a primary power-law spectrum. Each
panel shows the initial relic spectrum (blue), the total spectrum after the
acceleration event (orange), the directly accelerated spectrum
(green dashed line), and the reaccelerated spectrum (red dashed line). The
theoretically expected reaccelerated spectrum is also shown (black dashed line)
and matches the simulated reacceleration spectrum. The slope of the
reaccelerated spectrum in the weak-shock case deviates slightly from its
theoretical expectation because of numerical scatter of the shock compression
ratio (see Figure~\ref{fig:ShockTube1D_ShockStatistics}).

In the case of a strong shock, the primary accelerated spectrum dominates over
the reaccelerated spectrum, hence the total spectrum is only weakly modified by
reacceleration (see Figure~\ref{fig:ShockTube1D_Reacceleration_Comparison}). In
contrast, the reaccelerated spectrum dominates the total spectrum for large
momenta at weak shocks. This is important for observable signatures such as,
e.g. the flux of radio emission. We note that the relative strength between
direct acceleration and reacceleration depends on the details of shock
acceleration, which we do not resolve with our hydrodynamical simulations. In
our setup, we convert a fixed fraction of the accelerated CR proton energy into
CR electrons at the shock which leads to a larger normalisation for steeper
spectra. Other shock acceleration models, e.g. thermal leakage
models \citep{Kang2011}, predict different relative strengths of reacceleration
to direct acceleration.

\subsection{Sedov--Taylor blast wave}
\begin{figure*}
\includegraphics[width=\textwidth]{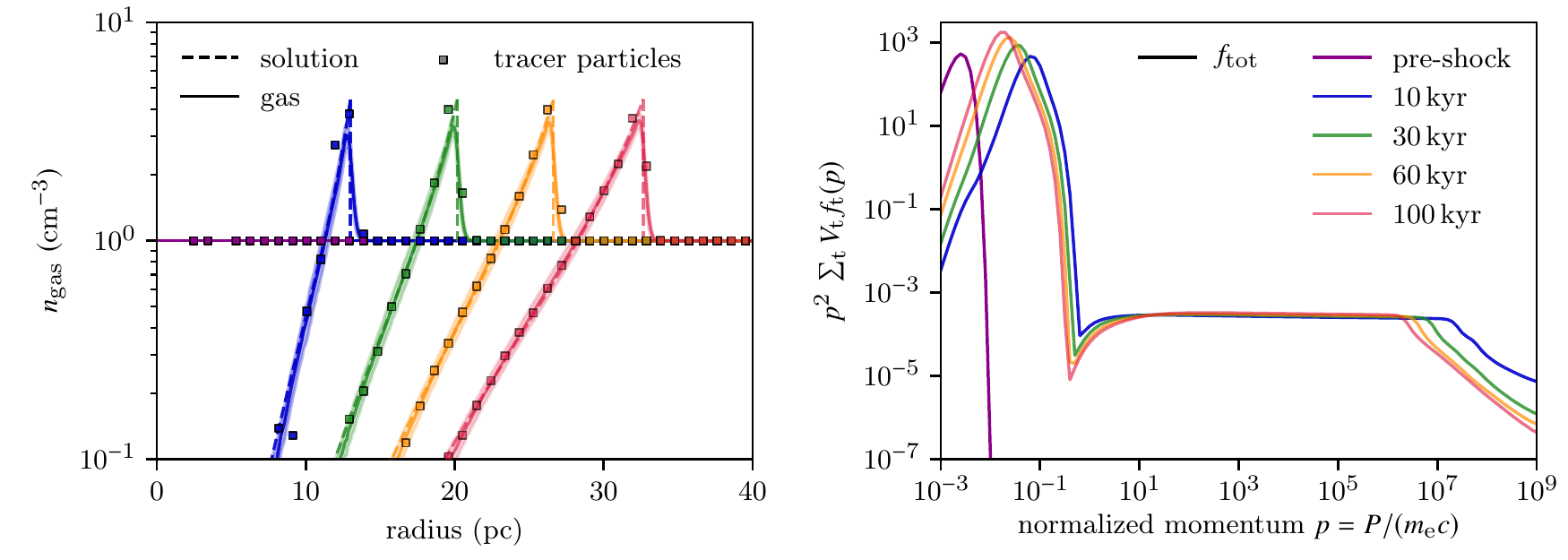}
\caption{3D Sedov--Taylor blast-wave simulation with $200^3$ gas cells
	and $\sim 30^3$ tracer particles. The left-hand panel shows the radial
	gas density profile (solid), the theoretical solution (dashed) and the
	spherically-averaged density of the tracer particles within concentric
	shells (points) for different times. The right-hand panel shows the
	total initial thermal spectrum (purple) and the total spectrum of
	particles, which have crossed the shock, at four characteristic times.}
\label{fig:Sedov_DensityProfile_OverviewPlot}
\end{figure*}

\begin{figure*}
\includegraphics[width=\textwidth]{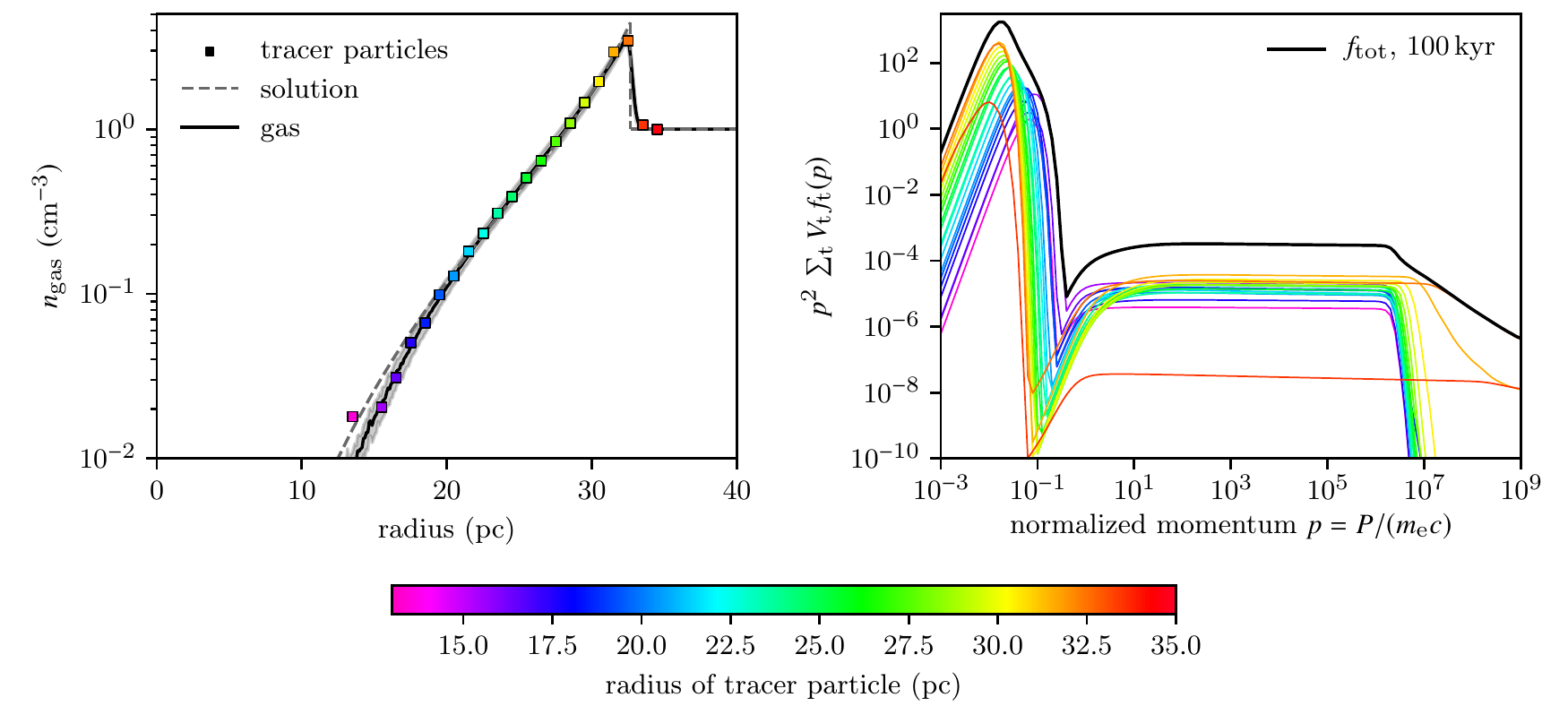}
\caption{Spherically-averaged tracer particle density (left-hand panel)
        and the contribution of tracer particles to the total spectrum
	(right-hand panel) for the 3D Sedov--Taylor blast-wave simulation
	at \SI{100}{kyr}. The colour indicates the different radii of tracer
	particles, which correspond to different times since shock crossing. The
	radial bins for the tracer particles have a width of \SI{1}{pc} and the
	grey band denotes the density scatter of the simulation.}
\label{fig:Sedov_ContributionToSpectrum}
\end{figure*}
In addition to the shock-tube tests we perform simulations of spherical shocks
in order to test acceleration and cooling in tandem with adiabatic CR electron
expansion. We setup a 3D Sedov--Taylor problem with an energy-driven spherical
shock which expands into a medium with negligible pressure.  We use a symmetric
3D box with $200^3$ cells, \SI{100}{\parsec} side length and the following parameters for the
initial conditions: The gas number density of the ambient medium is
$n_\mathrm{gas} = \SI{1}{\per\cubic\cm}$, has a temperature of $T=\SI{e4}{K}$
and a thermal adiabatic index of $\gamma_\mathrm{th} = 5/3$. We inject an initial thermal
energy of $E_0 = \SI{e51}{erg}$ into the central cell. The tracer particles are
initially located on a regular Cartesian mesh with $30^3$ grid points of which
we excise a small spherical region around the centre. The tracer particles initially only contain a thermal electron spectrum.

The left-hand panel of Figure~\ref{fig:Sedov_DensityProfile_OverviewPlot} shows
the simulated gas density profile for different snapshots together with the
theoretical solution and the spherically-averaged density of the tracer
particles within concentric shells. As expected for a single polytropic fluid,
the shock radius of the 3D explosion evolves as
\begin{equation}
r_{\rmn{shock}}(t) = \left(\frac{E_0}{\alpha \rho_0}\right)^{1/5} t^{2/5},
\end{equation}
where $\rho_0$ is the ambient mass density and $\alpha$ the self-similarity
parameter of the Sedov \citeyearpar{Sedov1959} solution.  In our simulation, we
adopt a CR shock acceleration efficiency of $\zeta_\mathrm{CR}=0.1$, which yields
an effective adiabatic index $\gamma_\mathrm{eff} = 1.58$ and a self-similarity
parameter $\alpha = 0.57$ \citep{Pais2018}. We note that the tracer particles
experience a slightly smaller density jump of $r\approx4$ in comparison to the
theoretically expected value of $r=4.45$ in this setup due to the narrow density
jump of the theoretical solution and the limited spatial resolution of the
hydrodynamical simulation.

The right-hand panel of Figure~\ref{fig:Sedov_DensityProfile_OverviewPlot} shows
the total electron spectrum, where we only take the spectrum of those particles
into account which have already crossed the shock front except for the initial
spectrum with all tracer particles. It is apparent that the total spectrum is
approximately constant for all snapshots at momenta $10^1 \lesssim p \lesssim
10^6$ where the
time-scales of Coulomb, inverse Compton, and synchrotron cooling are longer than
our simulation time. This is a consequence of constant kinetic and total energy
of the shock of a Sedov--Taylor blast wave. In the thin-shell approximation, all
mass is contained in a shell of radius $r_{\rmn{shock}}$ that expands with
velocity $\varv_{\rmn{post}}= 2\varv_{\rmn{shock}}/(\gamma_{\rmn{eff}}+1)$,
which yields a constant kinetic energy of $E_{\rmn{kin}}=32\pi E_0/\left[75 \alpha
(\gamma_\mathrm{eff} + 1)^2\right] \approx 0.35 E_0$. A fraction of this energy goes
into CR electrons, and particles that have recently crossed the shock dominate
the spectrum. We note that we obtain robust results for the CR electron spectrum
although the tracer particles are more coarsely sampled than the gas cells by a
factor of $\sim6^3$.

The contribution to the total spectrum of tracer particles at different radii is
shown in Figure~\ref{fig:Sedov_ContributionToSpectrum}. Red lines represent the
radial bin at which only a fraction of tracer particles experience shock
acceleration. Hence, their spectra are subdominant to the total spectrum. The
total spectrum is dominated by particles whose distance to the centre is close
the maximum of the radial gas density profile as they all experience shock
acceleration and have not yet lost energy due to adiabatic expansion. Yellow to
purple lines represent particles which are located towards inner radii and which
are effected by cooling due to adiabatic expansion and non-adiabatic processes.
These tests demonstrate that our code also handles adiabatic expansion together
with acceleration and cooling of CR electrons.  We note that the CR electron
spectrum is efficiently calculated with \textsc{crest}, which is faster than the
hydrodynamical simulation by a factor of about 460.

\section{Conclusions}
\label{sec:Conclusion}

We have presented our stand-alone post-processing code \textsc{crest} that
evolves the spectra of CR electrons on Lagrangian trajectories spatially and
temporally resolved. So far, we model the spatial CR electron transport as
advection with the gas and defer modelling CR electron streaming and (spatial)
diffusion to future work. All important physical cooling processes of CR
electrons are included, i.e. adiabatic expansion, Coulomb cooling, and
radiative processes such as inverse Compton, synchrotron and bremsstrahlung
cooling. In addition to adiabatic compression, we account for non-adiabatic
energy gain processes such as diffusive shock acceleration and reacceleration as
well as Fermi-II reacceleration via particle interactions with compressible
turbulence.

The CR electron cooling times at very low and very high momenta are much smaller than
typical time steps in simulations of galaxy formation or the ISM.  Hence, we develop a hybrid algorithm that combines numerical and
analytical solutions to the Fokker--Planck equations such that the resulting code
works efficiently and accurately on the MHD time step. We demonstrate in a
number of code validation simulations that the result of our hybrid algorithm
is as good as the fully numerical solution, which is however
computationally considerably more expensive.  This hybrid treatment decreases
the computational cost of evolving the CR electron spectrum and renders cosmological
simulations with CR electrons feasible.

\textsc{crest} has been extensively tested in idealized one-zone
models and alongside hydrodynamical simulations of the \textsc{arepo} code.
Idealized one-zone tests demonstrate that isolated terms of the Fokker--Planck
equation are accurately captured with our code. The \textsc{arepo} simulations
show (i) that \textsc{crest} works very well and efficiently together with a
hydrodynamical code at almost negligibly additional computational cost, (ii)
that the total spectrum, which is the sum of singular spectra on tracer
particles, evolves as expected, and (iii) that the spatial sampling of the
tracer particles quickly converges with increasing number of tracer
particles. In particular, our results are robust to a coarser sampling of the
tracer particles in comparison to the resolution of our unstructured
mesh. Future studies will show how the spectral properties depend on the spatial
sampling rate in more complex simulations of realistic environments.  We note
that our algorithm (and code) can in principle be combined with
every \mbox{(magneto)}hydrodynamical code that has Lagrangian tracer particles
on which the comoving Fokker--Planck equations for the CR electron spectrum is
solved.

The presented method allows studying the evolution of the CR electron spectrum
in the ISM, in galaxies and galaxy clusters as well as for AGN jets in great
detail.  It enables to link the non-thermal physics to observables such as
$\gamma$-ray and radio measurements and to distinguish leptonic and hadronic
emission scenarios. These include SNRs where we can gain insight which
environmental parameter (mean density, density fluctuations, magnetic field
strength) determines the dominating emission scenario. It will further allow us
to perform self-consistent studies on the evolution of the Fermi bubbles or
galactic outflows in MHD simulations and to test models that rely on star
formation or on AGN activity. This insight will be key for a more profound
understanding of the most important feedback processes during the formation of
galaxies. Finally, our code \textsc{crest} will enable us to self-consistently
follow the CR electron spectrum during the evolution of galaxy clusters, it can
possibly help to understand the enigmatic formation scenarios of radio relics
and radio haloes and how they relate to the dynamical state of clusters.

\section*{Acknowledgements}
It is a pleasure to thank Volker Springel for the use of \textsc{arepo}. We also thank the anonymous referee for constructive comments that helped to improve the paper.
We acknowledge support by the European Research Council under ERC-CoG grant CRAGSMAN-646955.




\bibliographystyle{mnras}
\bibliography{bibliography} 

%
%

\appendix

\section{Numerical solution to the Fokker--Planck equation}
\label{sec:AppA}
Here, we present details of the advection and diffusion operator. In the
following, $f_i^n = f(p_i, t_n)$ denotes the value of the spectrum at momentum
$p_i$ and time $t_n$.

\subsection{Advection operator}
\label{sec:advection_op}
Our advection operator accounts for cooling (i.e. Coulomb, bremsstrahlung,
inverse Compton, and synchrotron cooling), adiabatic changes, and particle
acceleration. In our code, we treat Fermi-I acceleration and reacceleration as
continuous injection via the term $Q(p,t)$ in equation~\eqref{eq:CRe_1D_FP_full}
as long as the tracer particle resides in shock surface and post-shock cells,
i.e. we treat CR electron acceleration identically to the model for CR proton acceleration
described by \citet{Pfrommer2017a}. The advection problem obeys the reduced
equation
\begin{align}
\left.\dv{f(p,t)}{t}\right|_\mathrm{adv} &- \pdv{p} \left\{ f(p,t)
\left[\frac{p}{3} \left(\div{\vb*{\varv}}\right)-\dot{p}(p,t)\right]\right\}=
\nonumber\\
&- \left(\div{\vb*{\varv}}\right) f(p,t) + Q(p,t),
\end{align}
where $\dv*{}{t} = \upartial/\upartial t + \bs{\varv\cdot\nabla}$ is the
Lagrangian time derivative.

We discretise this equation with a flux-conserving finite volume scheme using a
second-order piecewise linear reconstruction of the
spectrum \citep{LeVeque1988}. In addition, we use the non-linear van Leer flux
limiter \citep{Leer1977} and treat the terms on the right-hand side as an
inhomogeneity.

All following equations have in principle to be carried out for all $N$ momentum
bins, i.e. $i \in\left[0,\, N{-}1\right]$. The spectrum significantly decreases due to
rapidly cooling at low and high momenta. We thus cut the spectrum at
$f_\mathrm{cut}$ below which we treat numerical values of the spectrum as zero.
The related indices $i_\mathrm{lcut}$ and $i_\mathrm{hcut}$ are defined in
equations~\eqref{eq:Index_lcut} and \eqref{eq:Index_hcut}, respectively. The indices
$i_\mathrm{low}$ and $i_\mathrm{high}$ of the transition momenta between the
analytical and numerical solutions are defined in equations~\eqref{eq:Index_low}
and \eqref{eq:Index_high}. This further limits the momentum range of the
advection solver.  We define two limiting indices of the advection operator
\begin{align}
&i_\mathrm{ladv} = \max\mo\left(i_\mathrm{lcut}, i_\mathrm{low} \right) \mbox{ and}\\
&i_\mathrm{hadv} = \min\mo\left(i_\mathrm{hcut}, i_\mathrm{high} \right)
\end{align}
for which $0 \leq i_\mathrm{ladv}$ and $i_\mathrm{hadv} \leq N$ hold. In the case of a fully numerical simulation, the limiting indices of the advection operator are $i_\mathrm{ladv}=i_\mathrm{lcut}$ and $i_\mathrm{hadv} = i_\mathrm{hcut}$.

Because the total cooling time-scale $[\dot{p}(p) / p]^{-1}$ is a convex function
of momentum, the shortest cooling time-scale is determined by the smallest or
largest momentum of the momentum range which is treated by the advection
operator. As the numerical scheme
calculates fluxes between bins, we evaluate the maximum function over the
cooling rates in equation~\eqref{eq:TimeStep_Advection} on the outermost bin edges
and use
\begin{align}
\max\mo\left(\frac{\abs{\dot{p}(p)}}{\Delta p}\right) = \max\mo\left(\frac{\dot{p}\mo \left(p_{i_\mathrm{ladv} + 3/2}\right)}{p_{i_\mathrm{ladv} + 2} - p_{i_\mathrm{ladv} + 1}},\frac{\dot{p}\mo \left(p_{i_\mathrm{hadv}  - 5/2}\right)}{p_{i_\mathrm{hadv} - 1} - p_{i_\mathrm{hadv} -2}}   \right).
\end{align}
The advection operator has a symmetric stencil of five bins which makes in total
four ghost bins, with indices $i_\mathrm{ladv}$, $i_\mathrm{ladv}{+}1$,
$i_\mathrm{hadv}{-}2$, and $i_\mathrm{hadv}{-}1$, necessary.  The function values
on these bins are determined by power-law extrapolation.

The advection operator works as follows.  First, we explicitly evolve the
spectrum under the influence of (re)acceleration and injection by a half time step
\begin{align}
f_i^{n+1/2} = f_i^n + \frac{\Delta t}{2} Q_i^n \label{eq:Function_First_Half_Injection}
\end{align}
where $Q_i^n = Q(p_i, t_n)$ denotes the discretised (re)acceleration and injection rate at momentum
$p_i$ and time $t_n$ (see equations \eqref{eq:Fermi_I_reacceleration_source_function} and \eqref{eq:Fermi_I_acceleration_source_function}).  We define the advection velocity of momentum bin $p_i$ at
time $t_n$ due to adiabatic and cooling processes by
\begin{align}
u_i^n = \frac{p_i}{3} \left(\div{\vb*{\varv}}\right) - \dot{p}(p_i) \mbox{ at time } t_n.
\end{align}
The advection velocity of the bin edges $u_{i-1/2}$ is similarly defined.  We use
the advection velocities and the partly evolved function values $f_i^{n+1/2}$
from equation~\eqref{eq:Function_First_Half_Injection} to calculate fluxes $F$
through the bin edges at intermediate time $t_{n+1/2}$.  Depending on the sign
of the advection velocity $u_{i-1/2}$ at the bin edge, the flux is given by
\begin{align}
F_{i-1/2}^{n+1/2} = 
	\begin{aligned}[t]
	& u_{i-1} f_{i-1}^{n+1/2}\\
	& + \phi(r_{i-1}) \sigma_{i-1/2} \left(p_{i-1/2} - p_{i-1} + u_{i-1/2} \frac{\Delta t}{2} \right)
	\end{aligned}
\end{align}
for negative advection velocities $u_{i-1/2} < 0$ and by
\begin{align}
F_{i-1/2}^{n+1/2} =
	\begin{aligned}[t]
	& u_i f_i^{n+1/2}  \\
	& - \phi(r_{i}) \sigma_{i-1/2} \left(p_{i} - p_{i-1/2} - u_{i-1/2} \frac{\Delta t}{2} \right)
	\end{aligned} 
\end{align}
for positive advection velocities $u_{i-1/2} \geq 0$.
The variable $\sigma_i$ is the slope of the function values between two bins weighted with their advection velocities
\begin{align}
\sigma_{i - 1/2} =  \frac{u_i f_i^{n+1/2} - u_{i-1} f_{i-1}^{n+1/2}}{p_i - p_{i-1}}.
\end{align}
The function $\phi(r)$ is the slope limiter function, for which we use the van-Leer slope limiter
\begin{align}
\phi(r_i) = \frac{r_i + \abs{r_i}}{1 + \abs{r_i}}. \label{eq:vanLeer_SlopeLimiter}
\end{align}
The variable $r_i$ is ratio of slope at the left bin edge to the slope at the right bin edge, whose definition depends on the sign of the advection velocity.
For negative advection velocities $u_{i - {1/2}} < 0 $, we use
\begin{align}
r_{i-1} = \frac{u_{i-1} f_{i-1}^{n+{1/2}} - u_{i-2} f_{i-2}^{n+{1/2}} }{p_{i-1} - p_{i-2} } \frac{p_{i} - p_{i-1} }{u_{i} f_{i}^{n+{1/2}} - u_{i-1} f_{i-1}^{n+{1/2}} }
\end{align}
and for positive advection velocities $u_{i - {1/2}} \geq 0 $, we adopt
\begin{align}
r_{i} = \frac{u_{i+1} f_{i+1}^{n+{1/2}} - u_{i} f_{i}^{n+{1/2}} }{p_{i-1} - p_{i-2} } \frac{p_{i} - p_{i-1} }{u_{i} f_{i}^{n+{1/2}} - u_{i-1} f_{i-1}^{n+{1/2}} }.
\end{align}
We use the fluxes $F^{n+1/2}$ at intermediate time step $t_{n+1/2}$ with the half time step estimate of the spectrum $f^{n+1/2}$ to calculate the spectrum at time $t_{n+1}$
\begin{align}
f_i^{n+1} = 
\begin{aligned}[t]
& f_i^{n+{1/2}} \left( 1 - \Delta t \left(\div{\vb*{\varv}} \right)\right) + \frac{\Delta t}{2} Q_i\\
& + \Delta t \frac{F_{i+{1/2}} - F_{i-{1/2}}}{p_{i+{1/2}} - p_{i-{1/2}}} \left( 1 - \frac{\Delta t}{2} \left(\div{\vb*{\varv}} \right)\right),
\end{aligned}
\end{align}
where we included an additional factor $\left( 1 - \Delta t/2 \left(\div{\vb*{\varv}} \right)\right) $ that results from the influence of adiabatic changes on the fluxes.

\subsection{Diffusion operator}
\label{sec:diffusion_op}
The diffusion operator is based on the Crank--Nicolson method and solves the
diffusive part of the CR electron Fokker--Planck equation \eqref{eq:CRe_1D_FP_full},
\begin{equation}
\left.\dv{f(p,t)}{t}\right|_\mathrm{diff} + \pdv{p} \left[ \frac{f(p,t)}{p^2} \pdv{p}(p^2 D_{\!p\!p}) \right]
- \pdv[2]{p}\left[D_{\!p\!p} f(p,t)\right]=0.
\end{equation}
The combination of an explicit solution
\begin{align}
f_i^{n+1} = \alpha_i f_{i-1}^{n\phantom{+1}} + (1 - \beta_i) f_i^{n\phantom{+1}} + \gamma_i f_{i+1}^{n\phantom{+1}} \label{eq:general_explicit_solution}
\end{align}
and an implicit solution
\begin{align}
f_i^{n\phantom{+1}}	= -\alpha_i f_{i-1}^{n+1} + (1 + \beta_i) f_i^{n+1} - \gamma_i f_{i+1}^{n+1} \label{eq:general_implicit_solution}
\end{align}
yields the semi-implicit Crank--Nicolson scheme
\begin{align}
\begin{aligned}
&-\frac{\alpha_i}{2}f_{i-1}^{n+1} + \left( 1 + \frac{\beta_i}{2} \right) f_i^{n+1} - \frac{\gamma_i}{2} f_{i+1}^{n+1} \\
&\qquad= \frac{\alpha_i}{2}f_{i-1}^n + \left( 1 - \frac{\beta_i}{2} \right) f_i^n + \frac{\gamma_i}{2} f_{i+1}^{n}.
\end{aligned} \label{eq:general_crank_nicolson}
\end{align}
The coefficients $\alpha_i, \beta_i$, and $\gamma_i$ are derived by discretising the momentum diffusion term,
\begin{align}
\begin{aligned}
\alpha_i &= \left[ \frac{D_i}{p_{i+1/2} - p_{i-1/2}} + \frac{2 D_{i-1}}{p_{i-1}} \right] \frac{\Delta t}{p_i - p_{i-1}},\\
\beta_i &= \bigg[
	\begin{aligned}[t]
	& \frac{D_i}{p_i (p_i - p_{i-1})} + \frac{D_{i+1} - D_{i}}{(p_{i+1} - p_{i})^2} \\
	& + \frac{D_i}{p_{i+1/2}- p_{i-1/2}} \left( \frac{1}{p_{i+1} - p_i} + \frac{1}{p_{i} - p_{i-1}}\right) \bigg] \Delta t,
	\end{aligned} \\
\gamma_i &= \left[\frac{D_i}{(p_{i+1} - p_i)(p_{i+1/2} - p_{i-1/2})} + \frac{D_{i+1} - D_{i}}{(p_{i+1} - p_{i})^2} \right] \Delta t,
\end{aligned}
\end{align}
where we have used the abbreviation $D_i = D_{\!p\!p}(p_i)$ and the time step $\Delta t$. The diffusion time step is defined in equation \eqref{eq:TimeStep_Diffusion} which is
\begin{align}
\Delta t = \frac{C_\mathrm{CFL}}{D_0}
\end{align}
with $D_{\!p\!p} = D_0 p^2$ (see equation~\eqref{eq:MomentumDiffusion_Function}).
Equation~\eqref{eq:general_crank_nicolson} can be written as matrix equation
\begin{align}
\mat{A} \bs{\cdot} \vec{f}^{n+1} = \mat{B} \bs{\cdot} \vec{f}^{n} \label{eq:crank_nicolson_matrix_equation},
\end{align}
where $\mat{A},~\mat{B}$ are $(N\times N)$ matrices 
\begin{align}
\begin{aligned}
&\mat{A} =	\begin{pmatrix}
b_0		& 0			&			& 			& 			& 	\\
a_1		& b_1 		& c_1		& 	 		& 			& 	\\
		& a_2		& b_2		& c_2		& 			& 	\\
		& 			& \ddots	& \ddots	& \ddots	& 	\\
	 	&  			&  			& a_{N-2} 	& b_{N-2}	& c_{N-2}	\\
0		& 			& 			& 			& 0			& b_{N-1}	
\end{pmatrix}\\
&\mat{B} =	\begin{pmatrix}
1 			& 0				& 				& 					& 					& 		\\
\tilde{a}_1	& \tilde{b}_1 	& \tilde{c}_1	&  					& 					& 		\\
			& \tilde{a}_2	& \tilde{b}_2	& \tilde{c}_2		& 					& 		\\
			& 				& \ddots		& \ddots			& \ddots			& 		\\
		 	&  				&  				& \tilde{a}_{N-2}	& \tilde{b}_{N-2}	& \tilde{c}_{N-2}	\\
			& 				& 				& 					& 0					& 1	
\end{pmatrix}
\end{aligned}
\end{align}
and $\vec{f}^n$ and $\vec{f}^{n+1}$ are $N$-dimensional vectors containing the values of the CR electron spectrum in every momentum bin at time $t_n$ and $t_{n+1}$ respectively, i.e. $\vec{f}^n = (f_i^n)$ and $\vec{f}^{n+1} = (f_i^{n+1})$.
The coefficients of the matrices $\mat{A}$ and $\mat{B}$ are 
\begin{align}\
\begin{matrix}
a_i 		= -\frac{\alpha_i}{2},	& b_i 			= 1+\frac{\beta_i}{2},	& c_i 			= -\frac{\gamma_i}{2}, \\
\tilde{a}_i = \frac{\alpha_i}{2},		& \tilde{b}_i 	= 1-\frac{\beta_i}{2},	&\tilde{c}_i	= \frac{\gamma_i}{2},
\end{matrix}
\end{align}
for $i \in\left[1,\, N{-}2\right]$, whereas the coefficients $b_0^{} = f_0^n / f_0^{n+1}$ and $b_{N-1}^{} = f_{N-1}^n / f_{N-1}^{n+1}$ are chosen in order to fulfill the boundary conditions.
In order to solve equation~\eqref{eq:crank_nicolson_matrix_equation}, the tridiagonal matrix $\mat{A}$ is inverted with the Thomas algorithm, also known as tridiagonal matrix algorithm. If we write $\vec{d} = \mat{B} \bs{\cdot} \vec{f}^n$, the matrix equation~\eqref{eq:crank_nicolson_matrix_equation} takes the from
\begin{align}
\begin{pmatrix}
b_0 	& 0		&			&			&			&			\\
a_1		& b_1	& c_1		&			&			&			\\
& a_2	& b_2	& c_2 		&			&			\\ 
&		& \ddots&\ddots		& \ddots	&			\\
&		&		& a_{N-2}	& b_{N-2}	& c_{N-2} 	\\
&		&		&			& 0			& b_{N-1}
\end{pmatrix}
\begin{pmatrix}
f_0^{n+1}	\\ f_1^{n+1} \\ f_2^{n+1} \\ \vdots \\ f_{N-2}^{n+1} \\ f_{N-1}^{n+1}
\end{pmatrix}
=
\begin{pmatrix}
d_0	\\ d_1 \\ d_2 \\ \vdots \\ d_{N-2} \\ d_{N-1}
\end{pmatrix}
\end{align}
and its solution is numerically obtained by the application of the forward calculations
\begin{align}
c_i' = \left\{
\begin{array}{ll}
\frac{c_i}{b_i} & \mbox{for } i=0,\\
\frac{c_i}{b_i - a_i c_{i-1}'} & \mbox{for } i=1,\,\ldots,\,N{-}1 ,
\end{array}\right.\\
d_i' = \left\{
\begin{array}{ll}
\frac{d_i}{b_i} & \mbox{for } i=0,\\
\frac{d_i - a_i d_{i-1}'}{b_i - a_i c_{i-1}'} & \mbox{for } i=1,\,\ldots,\,N{-}1 ,
\end{array}\right.
\end{align}
and of the backward calculation
\begin{align}
f_i^{n+1} =  \left\{
\begin{array}{ll}
d_i' & \mbox{for } i=N{-}1,\\
d_i' - c_i' f_{i+1}^{n+1} & \mbox{for } i=N{-}2,\, \ldots,\,0 .
\end{array}\right.
\end{align}
We note that the amount of calculations can be reduced if we take only bins into
account where the spectrum is larger than a given low cut $f_\mathrm{cut}$. In
this case, equation~\eqref{eq:crank_nicolson_matrix_equation} reduces to an
$M$-dimensional matrix equation with $M = i_\mathrm{hcut} - i_\mathrm{lcut}$ and
the matrix inversion has to be applied for a submatrix, which is characterised
by the indices $i_\mathrm{lcut},\, i_\mathrm{lcut}{+}1,\, i_\mathrm{lcut}{+}2,\, \ldots,\, i_\mathrm{hcut}{-}2,\, i_\mathrm{hcut}{-}1$.


\bsp	
\label{lastpage}
\end{document}